\documentclass[aps,pra,twocolumn,superscriptaddress,showkeys]{revtex4-1}
\usepackage{graphicx}
\usepackage{subfigure}

\newcommand{\ket}[1] {| #1 \rangle}

\begin{document}

\title{Towards a room-temperature spin quantum bus in diamond via optical spin injection, transport and detection}

\author{M.W. Doherty}
\email{marcus.doherty@anu.edu.au}
\affiliation{Laser Physics Centre, Research School of Physics and Engineering, Australian National University, ACT 2601, Australia}

\author{C.A. Meriles}
\affiliation{Department of Physics, CUNY-City College of New York, New York, NY 10031, USA}

\author{A. Alkauskas}
\affiliation{Center for Physical Sciences and Technology, Vilnius LT-01108, Lithuania}

\author{H. Fedder}
\affiliation{3. Physikalisches Institut and Research Center SCOPE, University Stuttgart, Pfaffenwaldring 57, 70569 Stuttgart, Germany}

\author{M.J. Sellars}
\affiliation{Laser Physics Centre, Research School of Physics and Engineering, Australian National University, ACT 2601, Australia}
\affiliation{ARC Centre for Quantum Computation and Communication Technology, Australian National University, ACT 2601, Australia}

\author{N.B. Manson}
\affiliation{Laser Physics Centre, Research School of Physics and Engineering, Australian National University, ACT 2601, Australia}

\date{\today}

\begin{abstract}
Diamond is a proven solid-state platform for spin-based quantum technology. The nitrogen-vacancy (NV) center in diamond has been used to realize small-scale quantum information processing (QIP) and quantum sensing under ambient conditions. A major barrier in the development of large-scale QIP in diamond is the connection of NV spin registers by a quantum bus at room temperature. Given that diamond is expected to be an ideal spin transport material, the coherent transport of spin directly between the spin registers offers a potential solution. Yet, there has been no demonstration of spin transport in diamond due to difficulties in achieving spin injection and detection via conventional methods. Here, we exploit detailed knowledge of the paramagnetic defects in diamond to identify novel mechanisms to achieve spin injection, transport and detection in diamond at room temperature. Having identified these mechanisms, we explore how they may be combined to realise an on-chip spin quantum bus.
\end{abstract}


\maketitle

\section{Introduction}

The success of quantum information processing (QIP) in diamond is owed to the remarkable nitrogen-vacancy (NV) center \cite{review}. The spins of individual NV centers are optically addressable, such that light can be used to initialise and readout their electronic spin states \cite{review}. Additionally, the NV center has the longest spin coherence time ($T_2\sim2$ ms) of any solid-state system at room temperature \cite{balasubramanian09}. Using these properties, the room temperature operation of spin registers has been demonstrated in diamond by constructing clusters of NV centers or combining NV centers with other paramagnetic defects, such as $^{13}$C isotopic impurities and substitutional nitrogen N$_\mathrm{S}$ donor centers \cite{qip1,qip2,qip3,qip4,qip5,qip6,qip7,qip8}. At cryogenic temperatures, the coherence of the center's optical transitions has been exploited to entangle two spin registers separated by macroscopic distances via an optical quantum bus \cite{qip9,qip10,qip11}. However, since the coherence of the center's optical transitions is lost above $\sim$35 K, an optical quantum bus is not possible at room temperature \cite{fu09}. Thus, a major barrier in the development of scalable diamond QIP devices is the realization of a quantum bus between spin registers that is sufficiently compact to be incorporated on chip and can operate at room temperature \cite{awschalom13}.

One suggestion has been to establish a quantum bus by extending a chain of defect spins between two spin registers \cite{yao12,cap15}, but this is difficult to engineer and requires significant resources to control. An attractive alternative is the coherent electrical transport of a spin from one spin register to the other. Diamond presents as an ideal spin transport material: its large electronic bandgap, inversion symmetry, small spin-orbit interaction and low nuclear spin density promise long spin relaxation times \cite{restrepo12}. Indeed, \textit{ab initio} calculations predict a transport $T_1\sim$ 180 ns at room temperature \cite{restrepo12}, which corresponds to an exceptional transport distance of $\sim2$ mm in high purity diamond \cite{note1} with a modest electric field of $\sim100$ V/cm. Since only paramagnetic impurities (whose densities can be controlled) and magnetic field inhomogeniety offer additional dephasing mechanisms, transport is also expected to be coherent, with ultimately $T_2\sim T_1$ at room temperature \cite{balasubramanian09,jarmola12}.  Since coherent electrical spin transport can be controlled by nanoelectrodes \cite{nanoelectrodes}, which can be arranged in a scalable architecture \cite{hollenberg06}, it poses as a suitable quantum bus for scalable diamond QIP devices that operate at room temperature.

Despite this ample motivation, virtually no spin transport experiments in diamond have been reported. This palpable deficiency has been due to the apparent absence of any mechanism to inject spin into diamond \cite{zutic04}. Similar to silicon \cite{appelbaum07}, the indirect bandgap of diamond precludes conventional optical injection via circularly polarized light and conventional electrical injection is expected to be prevented by difficulties in interfacing ferromagnetic electrodes with diamond. Whilst advances in the interfacing of ferromagnetic electrodes ultimately enabled spin injection in silicon \cite{appelbaum07}, this method is not suitable for coherent transport between spin registers. Neither is the recent proposal to spin polarize carriers in conductive dopant wires in diamond heterostructures via Overhauser cross-relaxation with optically pumped ensembles of NV centers \cite{meriles14}.

\begin{figure}[hbtp]
\begin{center}
\includegraphics[width=0.9\columnwidth] {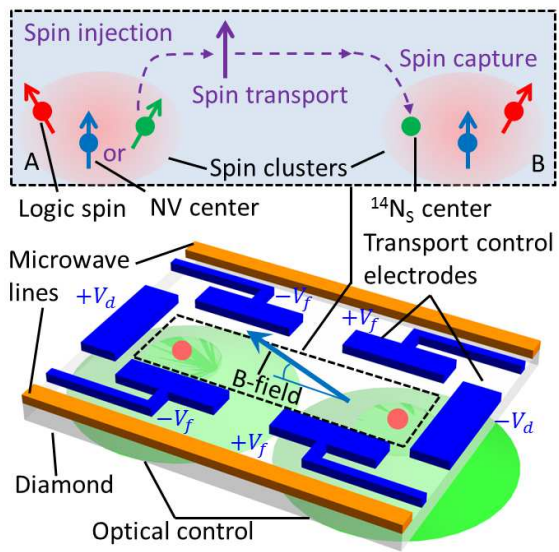}
\caption{(color online) (a) Example architecture of the proposed room-temperature spin quantum bus in diamond, including optically addressable spin clusters, an applied magnetic field (B-field) and surface microwave lines for spin control, and electrodes for spin transport control. Inset: detail of spin transport from cluster A to cluster B. Each cluster contains at least a NV center, a $^{14}$N$_\mathrm{S}$ donor center and a logic spin qubit (either the N nuclear spin of the NV center, a $^{13}$C isotopic impurity or another NV center). Optical spin injection occurs by using NV A to first prepare $^{14}$N$_\mathrm{S}$ A and then the photoionization of $^{14}$N$_\mathrm{S}$ A. Spin transport occurs as the ionized electron drifts through the diamond in response to the electric potential $V_d$ applied by the end electrodes. Additional focusing of the electron's drift is made by the potential $V_f$ applied by the lateral electrodes. Spin detection proceeds when the electron is captured by the ionized $^{14}$N$_\mathrm{S}$ B and its spin read using NV B.}
\label{fig:concept}
\end{center}
\end{figure}

In this paper, we explore how optical spin injection, transport and detection via defects in diamond may be combined to realise an on-chip spin quantum bus. We begin by reviewing the unique properties of the NV center and the paramagnetic donor centers: substitutional nitrogen N$_\mathrm{S}$ and phosphorus P$_\mathrm{S}$. In doing so, we identify how these properties permit novel optical spin injection and detection mechanisms. We ultimately conclude that NV-$^{14}$N$_\mathrm{S}$ defect pairs are the best option for coherent spin injection and detection. In subsequent sections, we combine the NV-$^{14}$N$_\mathrm{S}$ spin injection and detection mechanisms with a drift-diffusion model of electron transport to describe the coherent transport of spin between distant spin clusters. We then apply this model to propose a pathway to realizing an on-chip spin quantum bus for room-temperature diamond QIP (see Fig. \ref{fig:concept} for overview).

\section{The NV and donor centers}

\subsection{The NV center}

\begin{figure}[hbtp]
\begin{center}
\mbox{
\subfigure[]{\includegraphics[width=1\columnwidth] {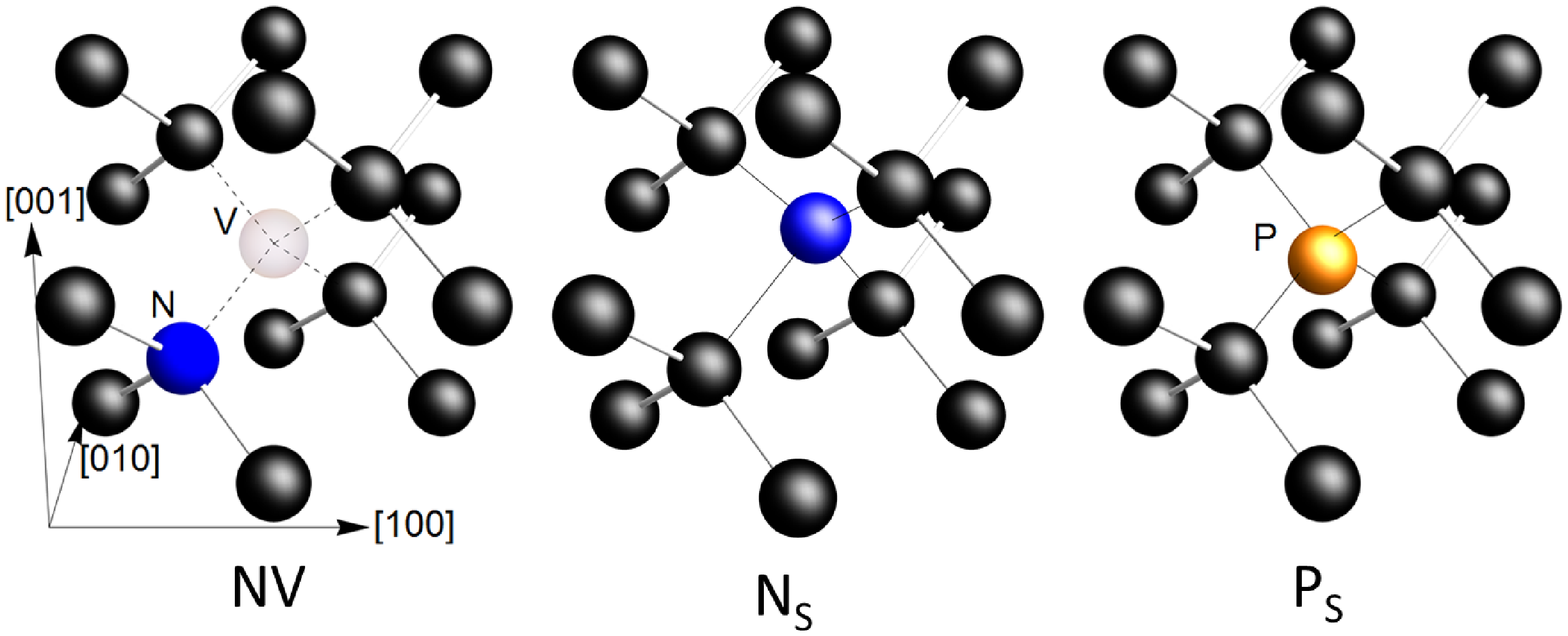}}
}
\mbox{
\subfigure[]{\includegraphics[width=0.75\columnwidth] {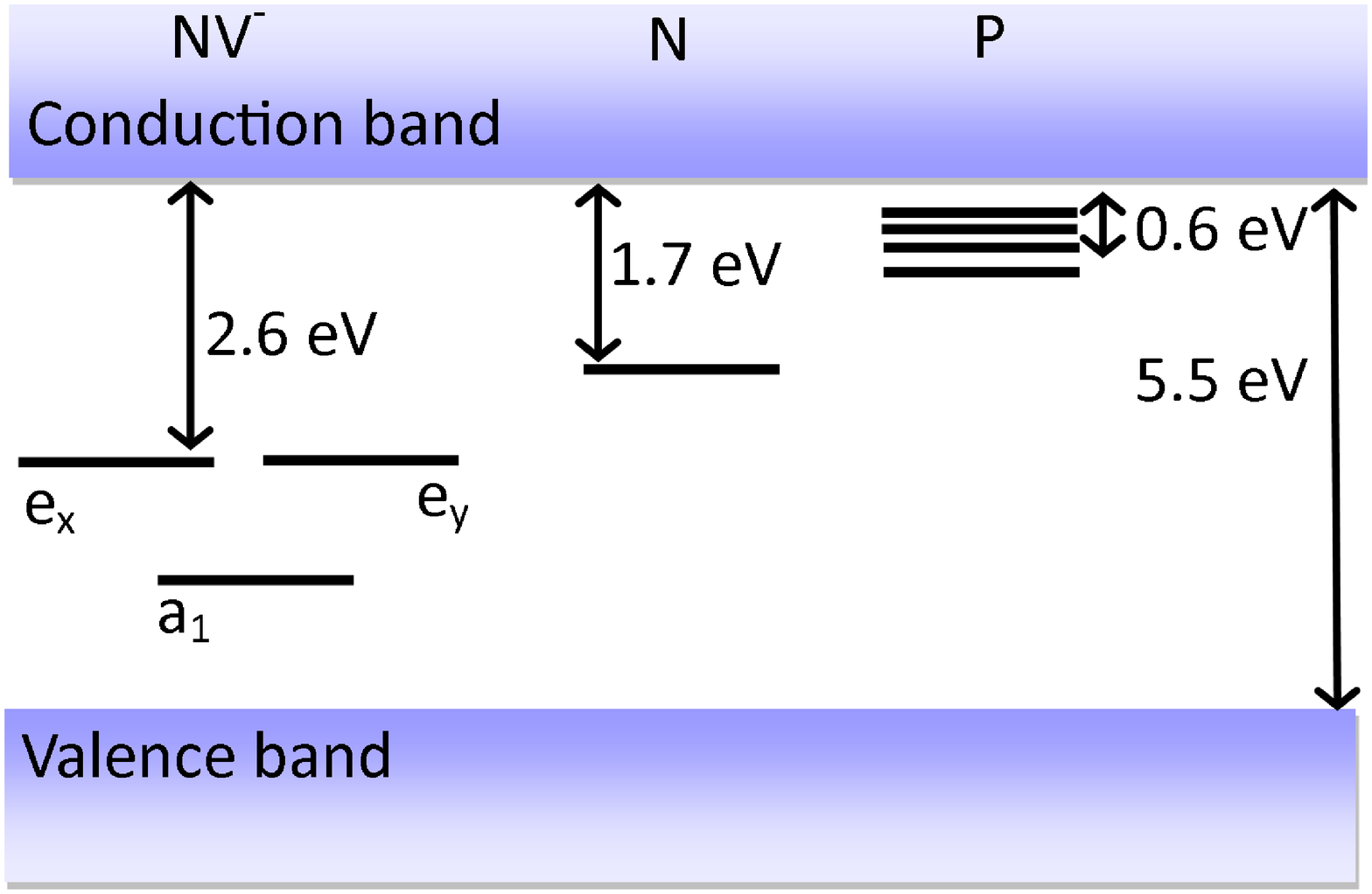}}
}
\mbox{
\subfigure[]{\includegraphics[width=0.7\columnwidth] {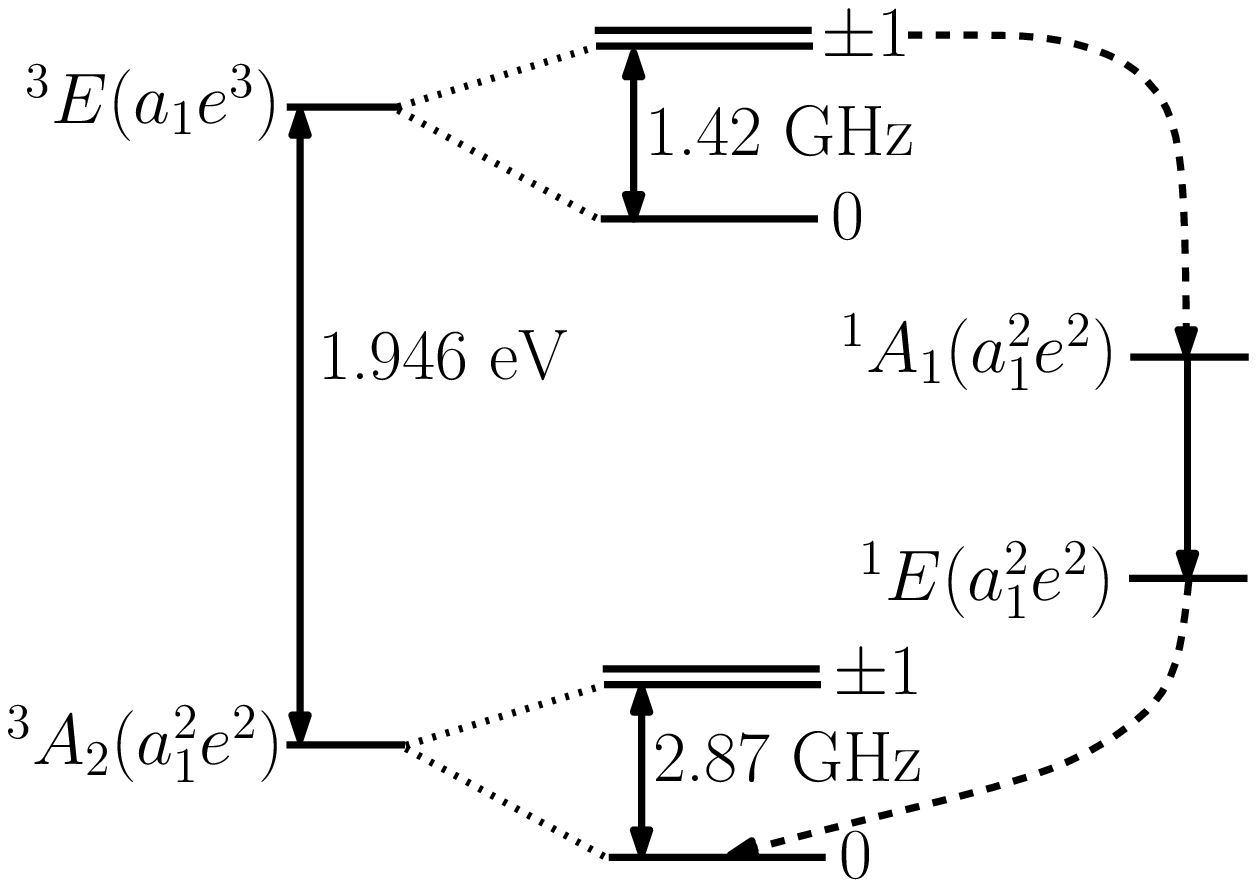}}
}
\caption{(color online) (a) diagrams of the NV, N$_\mathrm{S}$ and P$_\mathrm{S}$ centers, respectively, that depict carbon (black), nitrogen (blue) and phosphorus(orange) atoms and the vacancy (transparent). (b) electronic band structure diagram depicting the defect levels of the NV ($a_1$, $e_x$, $e_y$) and donor centers. The diamond indirect bandgap and defect photoionization energies are as indicated \cite{aslam13,nesladek98,katagiri06}. (c) electronic structure of the NV$^-$ center depicting the electronic levels with corresponding electronic configurations in parentheses, fine structure states (denoted by electronic spin projection) \cite{review}. The visible ZPL energy and the zero-field fine structure splittings of the triplet levels are as indicated. Solid arrows represent optical/ spin transitions, whereas dashed arrows represent the spin-selective non-radiative ISCs that lead to optical spin polarization and readout.}
\label{fig:defects}
\end{center}
\end{figure}

The NV center is a point defect comprised of a substitutional nitrogen-carbon vacancy pair orientated along the $\langle111\rangle$ crystallographic direction [see Fig. \ref{fig:defects}(a) for diagram and Ref. \onlinecite{review} for an extensive review of the NV center]. It is stable in neutral (NV$^0$) and negative (NV$^-$) charge states, which are characterised by their optical zero-phonon lines (ZPLs) at 2.156 eV (NV$^0$) and 1.946 eV (NV$^-$). NV$^-$ possesses the required properties for quantum technology and so is the desired charge state in thermal equilibrium. The thermal charge state is determined by the local density of electron donors and acceptors. N$_\mathrm{S}$ is the most prevalent donor in diamond and naturally coexists with the NV center as a by-product of NV formation: either through N CVD doping or ion implantation. Indeed, localized NV-N$_\mathrm{S}$ clusters, in which the N$_\mathrm{S}$ preferentially donates its electron to the NV to form NV$^-$, can be created by ion implantation \cite{qip1, pezzagna10}. It may be possible for NV-P$_\mathrm{S}$ clusters to be similarly created via N-P co-doping or co-implantation. However, we note that there are well known difficulties in the incorporation of phosphorus into diamond that reduce the yield of phosphorus atoms occupying substitutional, rather than split-vacancy, sites \cite{koizumi98,prins95,jones96}. The latter is problematic because it is a deep acceptor rather than a donor.

The observable electronic structures of NV$^0$ and NV$^-$ are defined by the occupation of the center's three deep-level defect orbitals ($a_1$, $e_x$, $e_y$) by three and four electrons, respectively [see Fig. \ref{fig:defects}(b)]. The electronic structure of NV$^-$ is depicted in Fig. \ref{fig:defects}(c) and consists of ground and optically excited spin triplet levels ($^3A_2$ and $^3E$) as well as intermediate spin singlet levels ($^1E$ and $^1A_1$). At room temperature, the observable fine structures of the $^3A_2$ and $^3E$ levels are analogous and can be described by the spin-Hamiltonian
\begin{eqnarray}
H_\mathrm{NV} & = & D(S_z^2-\frac{2}{3})+\gamma_eS_zB,
\label{eq:NVspinhamiltonian}
\end{eqnarray}
where $D$ is the zero-field splitting arising from spin-spin interaction that is equal to 2.87 GHz for $^3A_2$ and 1.42 GHz for $^3E$, $S_z$ is a $S=1$ dimensionless electron spin operator, $\gamma_e=g_e\mu_B/h$,  $\mu_B$ is the Bohr magneton, $g_e\sim2$ is the NV$^-$ electron g-factor that is approximately the isotropic free electron value, $h$ is the Planck constant and $B$ is the applied magnetic field that is aligned with the center's $\langle111\rangle$ trigonal symmetry axis. Note that hyperfine and interactions with electric and strain fields are typically of the order of $\sim1$ MHz and have not been included in (\ref{eq:NVspinhamiltonian}) because they do not significantly influence the discussion presented here.

Optical spin polarization and readout of NV$^-$ is enabled by spin-dependent non-radiative intersystem crossings (ISCs) between the spin triplet and singlet levels [depicted in Fig. \ref{fig:defects}(c)]\cite{goldman1,goldman2}. Upon optical excitation, the ISCs offer an additional decay pathway from $^3E$ to $^3A_2$ that preferentially depopulates $m_s=\pm1$ and populates $m_s=0$. After a few optical cycles the spin is polarized into $m_s=0$ with high fidelity. The additional non-radiative decay pathway is also responsible for the contrast in the center's fluorescence intensity between $m_s=0$ and $\pm1$. This contrast is used to optically read out the relative probability that the NV$^-$ spin is in the $m_s=0$ state. The time required to polarize/ readout the spin is defined by the lifetime $\sim300$ ns of the $^1E$ intermediate singlet level \cite{review}. With the addition of an appropriately tuned magnetic field, the NV$^-$ electron spin can also be used for high-fidelity projective initialization and readout of coupled nuclear spins \cite{qip8,qip12,qip13}

\begin{figure}[hbtp]
\begin{center}
\mbox{
\subfigure[]{\includegraphics[width=0.505\columnwidth] {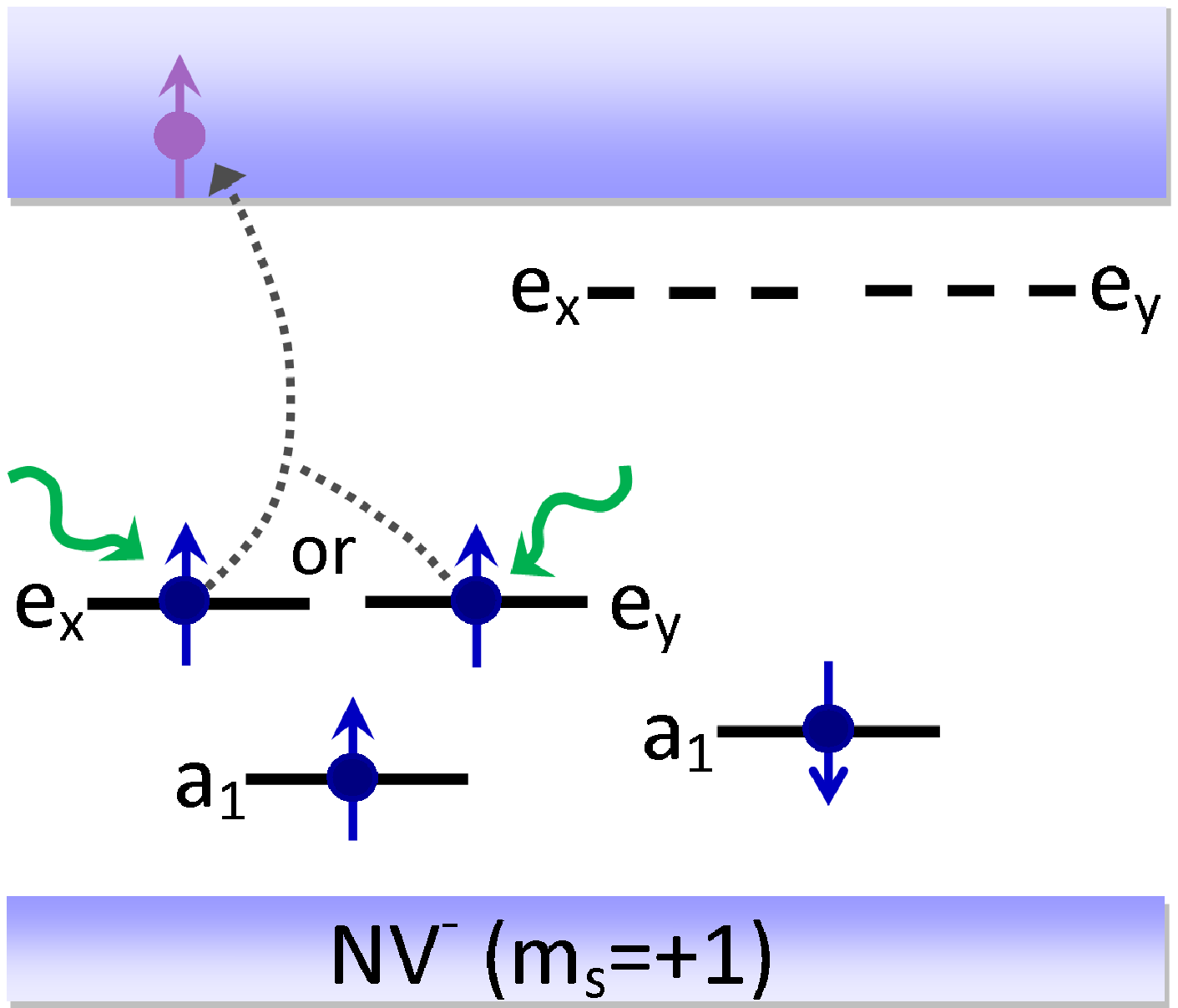}}
\subfigure[]{\includegraphics[width=0.495\columnwidth] {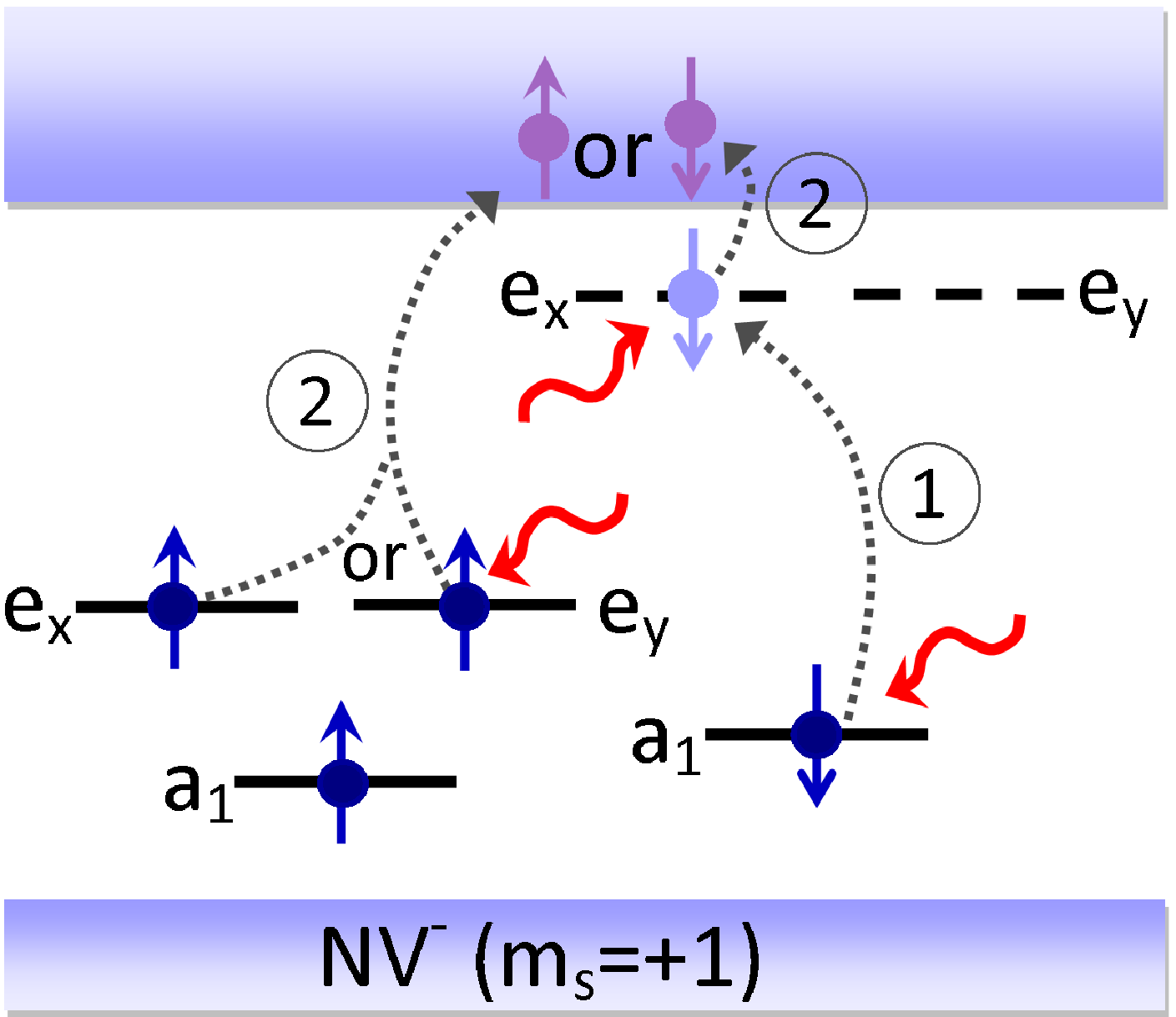}}
}
\mbox{
\subfigure[]{\includegraphics[width=1.0\columnwidth] {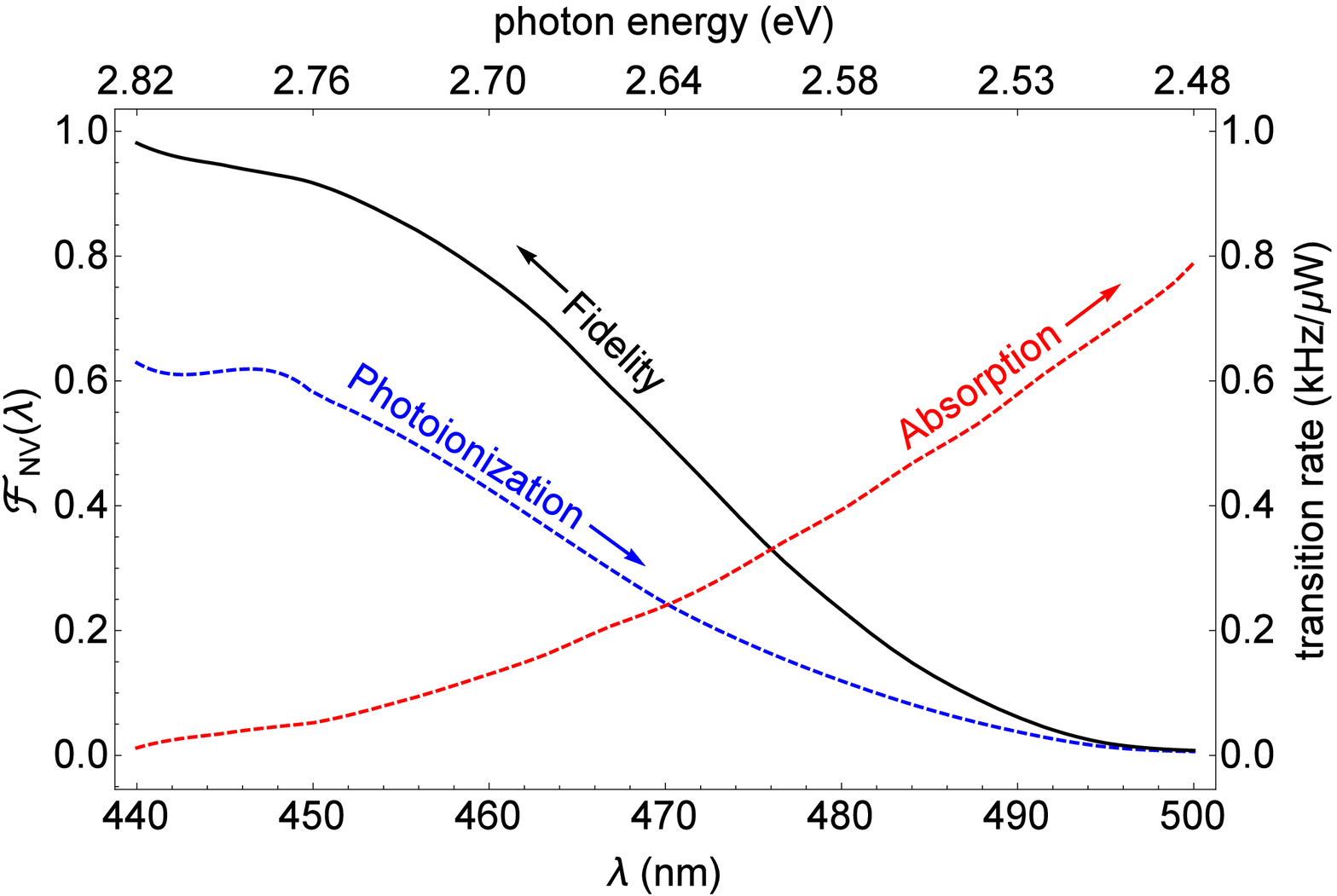}}
}
\caption{(color online) The one (a) and two (b) photon ionization mechanisms of NV$^-$. The depicted mechanisms correspond to photionization from the $m_s=+1$ spin state of the NV$^-$ ground $^3A_2$ level.  In (a), one photon with energy $>2.6$ eV ejects an electron from the spin-up $e$ spin-orbitals of NV$^-$ into the conduction band. In (b), one photon with energy $>1.946$ promotes an electron from the occupied spin-down $a_1$ spin-orbital to one of the unoccupied spin-down $e$ spin-orbitals (i.e. a transition from $^3A_2$ to $^3E$). A second photon with energy $>1.946$ eV ejects an electron from either the spin-up $e$ spin-orbitals or the occupied spin-down $e$ spin-orbital into the conduction band.
In (a-b): occupied/ unoccupied spin-orbitals are depicted as solid/ dashed lines; filled circles represent electrons and solid arrows denote spin-up/ -down electrons; wave-like arrows denote photons. (c) depicts the estimated ultimate spin-injection fidelity ${\cal F}_\mathrm{NV}(\lambda)$ for one-photon photoionization (solid black, left axis) \cite{estimatenote} as well as the one-photon photoionization \cite{aslam13} (dashed blue, right axis) and optical absorption \cite{davies74} (dashed red, right axis) transition rates per unit optical power (in the confocal arrangement described in Ref. \onlinecite{aslam13}).}
\label{fig:photoionisation}
\end{center}
\end{figure}

NV$^-$ can be photoionized into NV$^0$ by ejecting an electron into the diamond conduction band via the absorption of one photon with energy $>2.6$ eV \cite{aslam13} or by the successive absorption of two photons with energies $>1.946$ eV (see Fig. \ref{fig:photoionisation}) \cite{waldherr11}. In the latter mechanism, the first photon excites NV$^-$ to its optical excited state and the second photon ejects the electron into the conduction band. Detection of the photocurrent produced by NV$^-$ photoionization was reported recently \cite{bourgeois15}. NV$^0$ can be photoconverted back to NV$^-$ by a similar mechanism involving the ejection of a hole into the diamond valence band and the absorption of one photon with energy $>2.94$ eV or the successive absorption of two photons with energies $>2.156$ eV \cite{aslam13,waldherr11}. These back photoconversion mechanisms allow the desired negative charge state to be prepared on demand.

In greater detail, photoionization involves one of the NV center's electrons undergoing a transition from a defect spin-orbital to a conduction band Bloch spin-orbital $\phi_{n,\vec{k}}$, where $n$ denotes the band and $\vec{k}$ is the Bloch wavevector. If photons with energy close to the photoionization threshold energy are used, the transition will occur to a Bloch spin-orbital that belongs to the lowest conduction band and has a wavevector near the conduction band minimum. Diamond is a valley conductor with a conduction band minimum at the $\Delta\sim0.77\mathrm{X}$ k-point \cite{isberg13,saslow66}. Near the minimum, the lowest conduction band is non-degenerate and transforms as the $\Delta_1$ representation. This implies that there is no spin-orbit interaction near the conduction band minimum \cite{dresselhaus}. Thus, the photionization transitions near threshold will be spin-conserving because both the defect and Bloch spin-orbitals have the same well-defined spin. Band degeneracy, and therefore spin-orbit mixing, first occurs at the $X$-point, where the $X_1$ Bloch orbitals are located $\sim0.5$ eV above the conduction band minimum \cite{saslow66}. Thus, spin-conserving photoionization is primarily confined to photons with energies that are within $\sim0.5$ eV of threshold.

In the simplest picture, if a NV$^-$ center is optically prepared into the $m_s=+1$ spin state (or equivalently the $m_s=-1$ state) and then photoionized by the absorption of one photon with energy in the range $2.6-3.1$ eV, then a spin-up (-down) electron is ejected into the conduction band with 100\% probability. If the NV$^-$ center is instead photoionized by the successive absorption of two photons with energy $>1.946$ eV, then a spin-up (-down) electron is ejected into the conduction band with a much reduced 66\% probability because there is equal chance that the electron was ejected from any one of the three occupied (two spin-up and one spin-down) $e$ spin-orbitals in the $^3E$ optical excited state \cite{audnote}. However, this latter picture ignores the possibility that the NV$^-$ center may instead of absorbing a second photon and ionizing, decay back to the ground state. Owing to the optical spin-polarization mechanism, this decay may flip the NV$^-$ spin to the $m_s=0$ state, which reduces the probability that a spin-up(-down) electron is ejected during a subsequent photoionization event. It also ignores the possibility that absorption of the second photon from the optical excited state may ionize the electron 0.5 eV above threshold, where spin mixing may occur in the final conduction band orbitals, which will also reduce the probability. The lower 66\% probability plus these further reductions essentially eliminate the two-photon photoionization mechanism as an option for spin injection.

Noting that the differing spin-spin interactions in the NV$^-$ ground and optically excited states means that the electron spin dephases rapidly if the center is optically excited, the ultimate fidelity ${\cal F}_\mathrm{NV}$ for coherent spin-injection via one-photon photoionization is determined by the relative probability that, upon illumination, the center absorbs a photon and photoionizes or absorbs a photon and transitions to the optical excited state. The fidelity takes the simple expression
\begin{eqnarray}
{\cal F}_\mathrm{NV}(\lambda) = \frac{\sigma_\mathrm{ion.}(\lambda)}{\sigma_\mathrm{ion.}(\lambda)+\sigma_\mathrm{opt.}(\lambda)}
\end{eqnarray}
where $\lambda$ is the optical wavelength, and $\sigma_\mathrm{ion.}$ and $\sigma_\mathrm{opt.}$ are the wavelength dependent photoionization and optical absorption cross-sections, respectively. In Fig. \ref{fig:photoionisation}, we plot an estimate\cite{estimatenote} of ${\cal F}_\mathrm{NV}(\lambda)$ based upon data from Ref. \onlinecite{aslam13} for the one-photon photoionization rate $\propto\sigma_\mathrm{ion.}(\lambda)$ and the well-known NV$^-$ optical absorption spectrum $\propto\sigma_\mathrm{opt.}(\lambda)$ \cite{davies74}. As can be seen, $\sigma_\mathrm{opt.}(\lambda)$ decays towards shorter wavelength, whereas $\sigma_\mathrm{ion.}(\lambda)$ grows. ${\cal F}_\mathrm{NV}(\lambda)$ therefore increases towards shorter wavelength and, noting the limitations of our estimation, approaches unity near $\lambda\sim440$ nm (photon energy $2.8$ eV). Hence, we tentatively conclude that high spin-injection fidelity can be achieved via the one-photon photoionization mechanism. Furthermore, it appears that $\sigma_\mathrm{opt.}$ is such that this injection can occur rapidly over a timescale of $\sim1$ ns, if a $\sim1$ W optical pulse and a confocal arrangement are used.  Future work should seek to refine the measurements of $\sigma_\mathrm{ion.}(\lambda)$ and $\sigma_\mathrm{opt.}(\lambda)$ and our estimation of ${\cal F}_\mathrm{NV}(\lambda)$.

\subsection{The donor centers}

The N$_\mathrm{S}$ center is a deep donor in diamond and is known to exist in neutral (N$_\mathrm{S}^0$) and positive (N$_\mathrm{S}^+$) charge states. In N$_\mathrm{S}^0$, the N$_\mathrm{S}$ donor level is occupied by one electron, whereas in N$_\mathrm{S}^+$ it is unoccupied. There is disagreement about the energy of the N$_\mathrm{S}$ donor level below the conduction band, with various photoabsorption and photoconductivity measurements reporting the energy in the broad range 1.7-2.2 eV\cite{farrer69,heremans09,nesladek98,isberg06}. This disagreement is likely due to a weak onset of photoconduction, thereby making the measurement of the onset energy susceptible to detection sensitivity. Consequently, it is likely that the donor level exists towards the lower bound of the range, near 1.7 eV. Indeed, \textit{ab initio} calculations predict 1.8 eV \cite{deak14}. The P$_\mathrm{S}$ center is a comparatively shallow donor in diamond with six defect levels within 0.6 eV of the conduction band \cite{katagiri06,nesladek99,butorac08}. As per the N$_\mathrm{S}$ center, in the neutral charge state P$_\mathrm{S}^0$, one of these levels is occupied by an electron, whereas in the positive charge state P$_\mathrm{S}^+$, they are unoccupied \cite{katagiri06,nesladek99,butorac08}.

Both N$_\mathrm{S}^0$ and P$_\mathrm{S}^0$ photoionize via the absorption of one-photon with energy $>1.7$ eV and $>0.6$ eV, respectively, and the ejection of an electron into the conduction band. Thus, since photons with energy $>1.946$ eV are required to optically excite/ photoionize NV$^-$, the donor centers may be selectively photoionized using photons with energy $<1.946$ eV without affecting proximal NV$^-$. The photoconversion of N$_\mathrm{S}^+$ and P$_\mathrm{S}^+$ back to N$_\mathrm{S}^0$ and P$_\mathrm{S}^0$ requires UV photons with energies $>3.8$ eV and $4.9$ eV, respectively. Consequently, to achieve back conversion of a specific donor center to its neutral charge state, it may be more practical to instead photoionize a nearby ensemble of donor centers. The ejected electrons will diffuse and eventually be captured by the donor center of interest. For a diffusing electron density $\rho$ at the donor center, the capture rate will be $\Gamma_\mathrm{cap.}=\rho\sigma_\mathrm{cap.}\sqrt{k_BT/m}$, where $\sigma_\mathrm{cap.}$ is the capture cross-section of the donor center, $k_B$ is Boltzmann's constant, $T$ is temperature and $m$ is the effective electron mass \cite{alkauskas14}. The $\sigma_\mathrm{cap.}$ of N$_\mathrm{S}^+$ has been measured to be in the range $\sim3-7$ $\mathrm{nm}^2$ \cite{han95}, which implies that a reasonable $\Gamma_\mathrm{cap.}\sim1$ MHz is attained for a modest $\rho\sim5 \ \mathrm{\mu m}^{-3}$ at room temperature. To our knowledge, the electron capture cross-section of P$_\mathrm{S}^+$ has not been measured. Given that N$_\mathrm{S}$ is a deeper donor than P$_\mathrm{S}$, it is reasonable to expect that the electron capture cross-section of P$_\mathrm{S}^+$ is larger than that of N$_\mathrm{S}^+$ and will therefore achieve faster capture rates for the same diffusing electron densities.

From electron spin resonance (ESR) studies, it is known that N$_\mathrm{S}^0$ is distorted from the $T_d$ symmetry of the simple substitutional site to $C_{3v}$ symmetry by the elongation of one of the four N-C bonds, and is thus $\langle111\rangle$ orientated \cite{loubser78}. As is often mistaken, this distortion is not a Jahn-Teller effect because the ground electronic state has $A$ symmetry in the undistorted configuration \cite{stoneham92,cox94}. However, analogous to a Jahn-Teller distortion, N$_\mathrm{S}^0$ undergoes temperature dependent reorientation between its four equivalent distortions \cite{loubser67}. At room temperature, the reorientation rate is $\sim0.3$ kHz. It is similarly known from ESR studies that P$_\mathrm{S}^0$ is distorted to $D_{2d}$ symmetry with $\langle100\rangle$ orientation \cite{katagiri06}. Unlike N$_\mathrm{S}^0$, this is known to be due to the Jahn-Teller effect. The ESR broadens with temperature and eventually becomes undetectable at $\sim20$ K \cite{katagiri06,butorac08}. This broadening is most likely due to rapid reorientation via Jahn-Teller dynamics \cite{butorac08}. It is still an open question as to whether these dynamics reach the motionally-narrowed limit at higher temperatures and, as a consequence, if P$_\mathrm{S}^0$  has a usable electronic spin at room temperature. This question should be answered by future ESR or NV DEER measurements.

The N$_\mathrm{S}^0$ and P$_\mathrm{S}^0$ ESR do not exhibit zero-field fine structure from spin-orbit interactions and their electron g-factors are approximately isotropic and do not significantly differ from the free electron value \cite{katagiri06,loubser78}. Including hyperfine interactions, the spin-Hamiltonian that describes the ESR of the dopant centers is
\begin{eqnarray}
H_\mathrm{d} & = & \gamma_es_zB+\vec{s}\cdot\textbf{A}\cdot\vec{I}+\vec{I}\cdot\textbf{Q}\cdot\vec{I}+\gamma_nI_zB,
\label{eq:dopantspinhamiltonian}
\end{eqnarray}
where $\vec{s}$ are the $S=1/2$ dimensionless electron spin operators, $\vec{I}$ are the dimensionless nuclear spin operators, $\textbf{A}$ and $\textbf{Q}$ are the magnetic and electric hyperfine tensors, respectively, that depend upon the orientation of the donor center and the principal magnetic ($A_\parallel$, $A_\perp$) and electric ($Q$) hyperfine parameters, $\gamma_n=g_n\mu_N/h$, $g_n$ is the nuclear g-factor and $\mu_N$ is the nuclear magneton. Note that the spin quantization axes in (3) have been chosen to be parallel with the applied magnetic field $B$.

\begin{table}
\caption{\label{tab:trapsandcenters} Spin parameters of N$_\mathrm{S}^0$ and P$_\mathrm{S}^0$ taken from Refs. \onlinecite{katagiri06} and \onlinecite{cox94}: orientation of defect structure/ principal hyperfine axis, electronic g-factor, nuclear spin of isotope, and magnetic ($A_\parallel$, $A_\perp$) and electric ($Q$) hyperfine parameters.}
\begin{ruledtabular}
\begin{tabular}{llllllll}
Dop. &  Ori. & $g_e$ & $g_n$ & $I$ & $A_\parallel$ & $A_\perp$ & $Q$ \\
  &  & & & & (MHz) & (MHz) & (MHz)  \\
\hline
$^{14}$N$_\mathrm{S}^0$ &  $\langle111\rangle$ & 2  & +0.40 & 1 & 114 & 81  & -3.97 \\
$^{15}$N$_\mathrm{S}^0$ &  $\langle111\rangle$ & 2  & -0.57 & 1/2 & -160 & -114  & - \\
P$_\mathrm{S}^0$ &  $\langle100\rangle$ & 2  &  +2.26 & 1/2 & 162 & 33.9 & - \\
\end{tabular}
\end{ruledtabular}
\end{table}

Unlike NV$^-$, there are no competing optical processes that may depolarize/ dephase the donor electron spin when it is photoionized. However, as can be seen in Table \ref{tab:trapsandcenters}, the $^{15}$N$_\mathrm{S}^0$ and P$_\mathrm{S}^0$ hyperfine parameters are of the magnitude $\sim10^2$ MHz, which poses a major problem to coherent optical spin injection and detection using centers with these isotopes. Owing to the stochastic nature of photoionization, the time the donor ionizes is not precisely known. Due to the uncertainty in this ionization time, the donor electron spin acquires an unknown phase through hyperfine interaction with the $I=1/2$ nuclear spins of these isotopes. The same is true when the electron is captured: the electron spin gains an unknown phase through hyperfine interactions at the new center because the capture time is not precisely known. Thus, these hyperfine interactions will dephase the electron spin on a timescale of $\sim 10$ ns during photoionization and capture.

This dephasing can only be avoided by engineering photoionization and capture to occur on much shorter timescales (i.e. $\ll10$ ns). Whilst the photoionization cross-section of N$_\mathrm{S}^0$ is not precisely known, it is known for P$_\mathrm{S}^0$ to be $\sim0.75$ $\mathrm{\AA}^2$ \cite{pajot}. We may assume that it will be similar for N$_\mathrm{S}^0$. Therefore, to achieve a photoionization time that is $<1$ ns, the light power applied to a diffraction limited spot (for wavelength $\sim640$ nm) must be $>10$ mW, which is achievable. However, it is much more difficult to imagine how the capture time can be reduced to this timescale. This would require the transport of the injected electron to be controlled to the degree that its probability density at the capturing center is $>5\times10^3 \ \mathrm{\mu m}^{-3}$, which roughly corresponds to confining the electron to a cube with sidelengths $\sim60$ nm. As will be discussed further in section IV, this presents a major technical challenge, and upon this basis, we rule out  $^{15}$N$_\mathrm{S}^0$ and P$_\mathrm{S}^0$ as being practically viable options for coherent electron spin injection. We wish to note that they can, however, be used to simply inject incoherent spin-polarized electrons.

The situation is different for $^{14}$N$_\mathrm{S}^0$ because the dephasing due to hyperfine interactions can be avoided by preparing the $I=1$ spin of $^{14}$N$_\mathrm{S}^0$ into the $m_I=0$ nuclear spin projection, which has no magnetic hyperfine interaction with the electron spin. Several techniques for this preparation have been devised and demonstrated \cite{qip1,qip2,qip6,qip7}. Each involve optically spin-polarizing a proximal NV$^-$ and transferring this polarization to the nuclear spin of the $^{14}$N$_\mathrm{S}^0$ center via the dipolar coupling of the NV$^-$ and $^{14}$N$_\mathrm{S}^0$ electron spins. Given this preparation, the electron spin dephasing rate is reduced to the rate at which the nuclear spin flips to $m_I=\pm1$ and returns hyperfine interactions. The nuclear spin flip rate is determined by the non-secular terms in (\ref{eq:dopantspinhamiltonian}), which exist if the orientation of the magnetic field (aligned with the axis of the proximal NV$^-$) differs from the orientation of the $^{14}$N$_\mathrm{S}^0$ center. Due to the continuous (slow) reorientation of the N$_\mathrm{S}^0$ center at room temperature, differences in orientation are unavoidable. The effect of the non-secular terms can be minimized by the application of a sufficiently large magnetic field (e.g. $B\sim5000 \ \textrm{G}\gg A_\perp/\gamma_e\sim50$ G), such that non-secular terms between electron spin projections can be ignored. In which case, (\ref{eq:dopantspinhamiltonian}) becomes for $^{14}$N$_\mathrm{S}$
\begin{eqnarray}
H_\mathrm{d}& \approx & \gamma_es_zB+\chi A_\parallel s_zI_z+Q\left[I_z^2\cos^2\alpha+I_x^2\sin^2\alpha-\frac{2}{3}\right] \nonumber \\
&&+\gamma_nBI_z\cos\alpha-\frac{Q}{2}(I_xI_z+I_zI_x)\sin2\alpha \nonumber \\
&& +\gamma_nBI_x\sin\alpha
\label{eq:dopantspinhamiltoniansecular}
\end{eqnarray}
where $\chi=1$ if the orientation of the $^{14}$N$_\mathrm{S}^0$ center is aligned with the magnetic field and $\chi\approx\sqrt{5}/3$ if not, the nuclear spin quantization axis has been rotated by an angle $\alpha$ from the magnetic field $B$ that is defined by $\tan\alpha=0$ or $\tan\alpha=-(A_\parallel-A_\perp)\sin2\theta/[A_\parallel+A_\perp+(A_\parallel-A_\perp)\cos2\theta]$ for aligned/ not-aligned, respectively, and $\theta=\arccos(-1/3)$ is the tetrahedral angle. The final two terms drive the nuclear spin flips and are of the order of $\sim1$ MHz.

This remaining dephasing rate defines the maximum allowable period ($\sim1 \ \mathrm{\mu s}$) between preparation of the $^{14}$N$_\mathrm{S}^0$ nuclear spin and injection/ detection. During this period, two-qubit gates must be applied to the NV$^-$ and $^{14}$N$_\mathrm{S}^0$ electron spins in order to prepare/ detect the $^{14}$N$_\mathrm{S}^0$ electron spin state. It is therefore an important remaining question as to whether these gates can be performed sufficiently fast. Their speed is limited by the dipolar coupling between the electron spins, which is a function of the distance between them. Since NV$^-$-$^{14}$N$_\mathrm{S}^0$ pairs with separations as small as 1.5 nm and dipolar couplings as large as 14 MHz have been engineered by ion-implantation in the past, it appears that this speed requirement can be met with adequate engineering. Hence, we conclude that $^{14}$N$_\mathrm{S}$ centers are suitable for coherent spin injection and detection.

\subsection{Conclusions}

To summarize the key conclusions of this section, we have identified that:

(A) Localized NV$^-$-donor center clusters can be fabricated with dipolar couplings of the order $\sim10$ MHz.

(B) The NV$^-$ charge state and spin can be optically prepared.

(C) High-fidelity coherent spin injection can be achieved when a NV$^-$ photoionizes by the absorption of one photon with energy in the range $\sim$2.8-3.1 eV.

(D) The positive/ neutral charge states of donor centers can be prepared by selective photoionization/ capture of electrons photoionized from a nearby ensemble of donor centers.

(E) Owing to hyperfine interactions, the $^{15}$N$_\mathrm{S}$ and P$_\mathrm{S}$ centers are unsuitable for coherent spin injection, but may be used to inject spin-polarized electrons.

(F) By preparing the nuclear spin of the $^{14}$N$_\mathrm{S}$ center, it can be used to achieve high-fidelity coherent spin-injection/ detection without influencing nearby NV$^-$ as long as photons with energies in the range $\sim1.7-1.946$ eV are used, a magnetic field $B\sim5000$ G is applied and injection/ detection occurs over a period $<1 \ \mathrm{\mu s}$.

\section{Optical spin injection and detection}
In this section, we apply the conclusions of the previous section to describe how an electron with a well defined spin state $\ket{\psi}=\alpha\ket{\uparrow}+\beta\ket{\downarrow}$ can be optically injected into the diamond conduction band via either a single NV center or a NV-$^{14}$N$_\mathrm{S}$ pair and subsequently captured and optically detected by a NV-$^{14}$N$_\mathrm{S}$ pair.

\subsection{Injection via a NV center}

As depicted in Fig. \ref{fig:injectiondetection}(a), three steps are required to inject an electron with spin state $\ket{\psi}$ using a NV$^-$ center: (1) initialization, (2) preparation and (3) photoionization. In (1), the NV center is optically prepared into NV$^-$ and spin-polarized into the $\ket{0}=\sqrt{1/2}(\ket{\uparrow}\ket{\downarrow}+\ket{\downarrow}\ket{\uparrow})$ triplet spin state via a $\sim500$ ns green optical pulse. Here, on the righthand side, we have explicitly written the spin states of the two $e$-orbital electrons that  form the triplet. Now, when the NV$^-$ is photoionized, either of these electrons can be injected and so both must be prepared in $\ket{\psi}$. In (2), this is done by applying $\sim 10$ ns microwave control pulses to form the superposition of spin triplet states: $\ket{\Psi}=\alpha^2\ket{+1}+\sqrt{2}\alpha\beta\ket{0}+\beta^2\ket{-1}=(\alpha\ket{\uparrow}+\beta\ket{\downarrow})(\alpha\ket{\uparrow}+\beta\ket{\downarrow})$, which is the desired product state of the two electrons in $\ket{\psi}$. In this product state, separate spin-Hamiltonians for the two electrons can be defined. They are identical and equal to
\begin{eqnarray}
H_\mathrm{NV}^\mathrm{sep.} = \gamma_e s_z B +D(|\alpha|^2-|\beta|^2)s_z
\label{eq:NVhamiltoniansep}
\end{eqnarray}
Thus, similar to our previous discussion of hyperfine induced dephasing during injection, unless $\ket{\psi}$ is restricted to the subspace $|\alpha|^2=|\beta|^2$ (i.e. vanishing net dipolar interaction between the two spins), the electron spins will dephase on a timescale of $\sim1/D\sim0.35$ ns during injection. Consequently, to minimize dephasing during injection, a greater than $1$ W blue (2.8-3.1 eV) photoionization pulse must be used in step (3). Such high optical powers are likely to be unfeasible in scalable devices and thus, we rule out coherent spin injection via NV centers for such devices, unless they can operate in the restricted $|\alpha|^2=|\beta|^2$ subspace.

\subsection{Injection via a NV-$^{14}$N$_\mathrm{S}$ pair}

\begin{figure}[hbtp]
\begin{center}
\mbox{
\subfigure[]{\includegraphics[width=1.0\columnwidth] {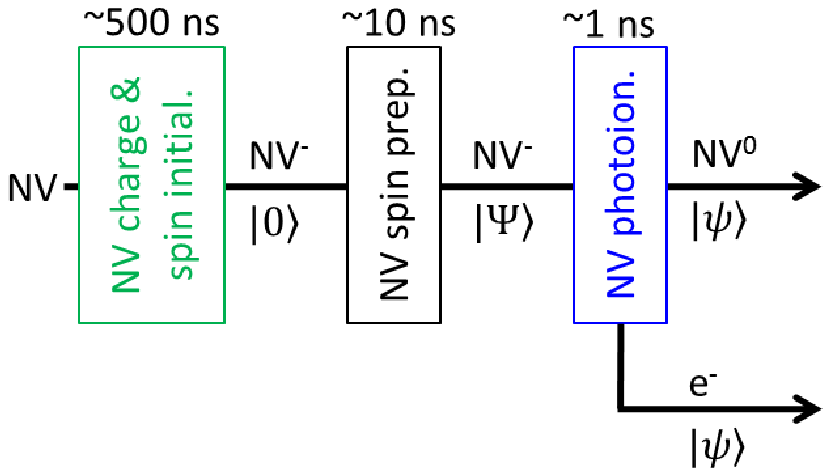}}
}
\mbox{
\subfigure[]{\includegraphics[width=1.0\columnwidth] {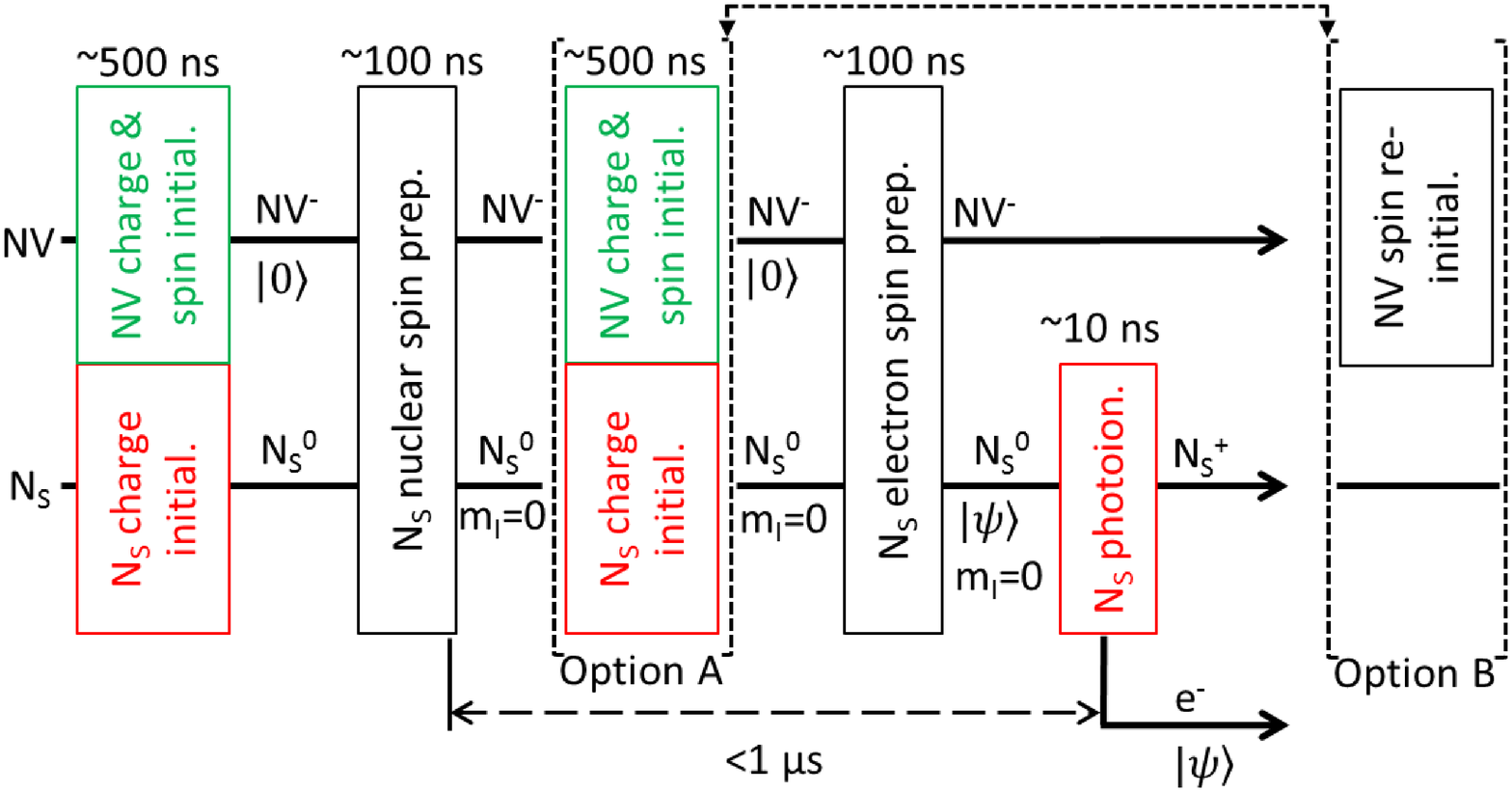}}
}
\mbox{
\subfigure[]{\includegraphics[width=1.0\columnwidth] {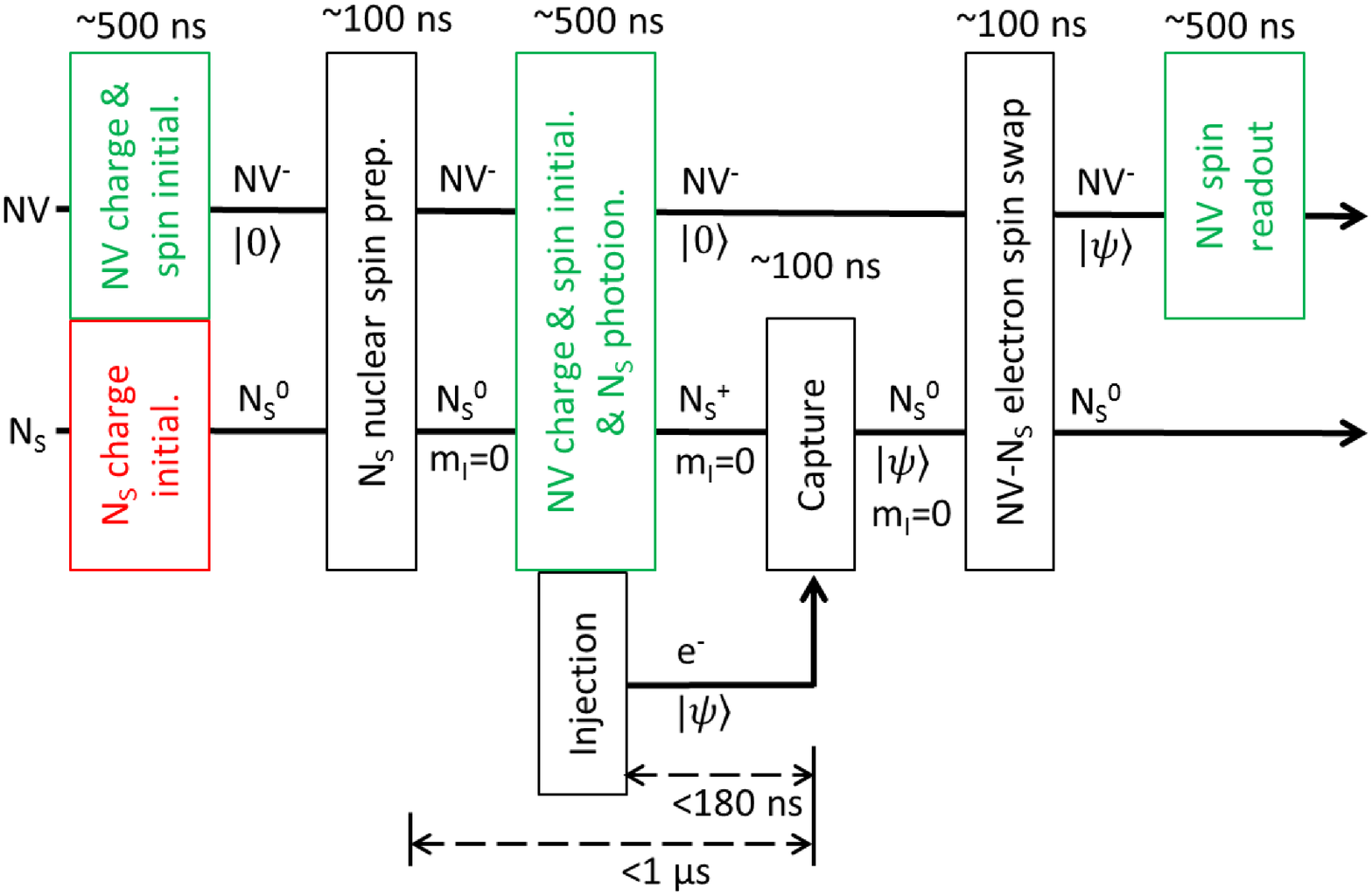}}
}
\caption{(color online) Techniques to optically inject an electron with spin state $\ket{\psi}$ via a NV center (a) or via a NV-$^{14}$N$_\mathrm{S}$ pair (b) and to subsequently optically detect it via a NV-$^{14}$N$_\mathrm{S}$ pair (c). Each box corresponds to an optical/ microwave pulse described in the text and has its indicative length denoted above. For optical pulses, the box color represents the wavelength of the pulse (red:1.7-1.946 eV, green:typically 2.32 eV, blue:2.8-3.1 eV). The evolution of the charge and spin states of each defect and the injected electron are represented by solid arrows and are as annotated. The respective $1 \ \mathrm{\mu s}$ and $180$ ns time frames for $^{14}$N$_\mathrm{S}$ nuclear spin and injected electron spin relaxation are indicated by dashed arrows.}
\label{fig:injectiondetection}
\end{center}
\end{figure}

As depicted in Fig. \ref{fig:injectiondetection}(b), three steps are required to inject an electron with spin state $\ket{\psi}$ using a NV$^-$-$^{14}$N$_\mathrm{S}$ pair: (1) initialization of the NV$^-$ spin and the charge and nuclear spin state of the $^{14}$N$_\mathrm{S}$, (2) preparation of the $^{14}$N$_\mathrm{S}$ electron spin into $\ket{\psi}$, and (3) photoionization of $^{14}$N$_\mathrm{S}$. In (1), the NV$^-$ electron spin is first spin-polarized by a green optical pulse and the $^{14}$N$_\mathrm{S}$ is prepared into the neutral charge state by capturing an electron photoionized from a nearby ensemble of donor centers by a red optical pulse. The NV$^-$ electron spin is then used to prepare the $^{14}$N$_\mathrm{S}$ nuclear spin into $m_I=0$ via microwave pulses \cite{qip1,qip2,qip6,qip7} before being re-polarized by another green optical pulse. During this second green pulse, the $^{14}$N$_\mathrm{S}$ may photoionize, so a second red pulse and diffusion-capture process must occur to re-prepare $^{14}$N$_\mathrm{S}$ in the neutral state. Alternatively, if the NV$^-$ nuclear spin was also initialized during the first initialization pulse using a projective technique \cite{qip8,qip12,qip13}, then the nuclear spin and microwave pulses can be later used to re-initialize the NV$^-$ electron spin rather than a second optical pulse. The advantage of this alternative is that there is no interference with the $^{14}$N$_\mathrm{S}$ and so, no requirement for charge re-initialization. In (2), the NV$^-$ electron spin and microwave pulses are used to prepare the $^{14}$N$_\mathrm{S}$ electron spin in $\ket{\psi}$. Finally, in (3) a red optical pulse is applied to photoionize the $^{14}$N$_\mathrm{S}$ and inject the prepared electron spin. Given the hyperfine dephasing of the $^{14}$N$_\mathrm{S}$ electron spin, steps (2) and (3) must be completed within $\sim1 \ \mathrm{\mu s}$ of the preparation of the $^{14}$N$_\mathrm{S}$ nuclear spin in (1). This can be comfortably achieved if the NV$^-$-$^{14}$N$_\mathrm{S}$ dipolar coupling is $\sim10$ MHz, so that the NV$^-$-$^{14}$N$_\mathrm{S}$ two-qubit gates that are used in (2) to prepare the $^{14}$N$_\mathrm{S}$  electron spin are limited to $\sim100$ ns.

\subsection{Detection via a NV-$^{14}$N$_\mathrm{S}$ pair}

There are three steps to optical detection [see Fig. \ref{fig:injectiondetection}(c)]: (1) initialization of the NV$^-$ spin and the charge and nuclear spin state of the $^{14}$N$_\mathrm{S}$, (2) capture of the injected electron by the $^{14}$N$_\mathrm{S}$, and (3) optical readout of the $^{14}$N$_\mathrm{S}$ electron spin via the NV$^-$ spin. Step (1) is as per injection, except the $^{14}$N$_\mathrm{S}$ is not re-charged at the end and left in the positive charge state. In (2), there is first a wait time for the electron to be injected and transported from elsewhere in the diamond and then the injected electron is captured. In (3), the $^{14}$N$_\mathrm{S}$ electron spin is readout by the NV$^-$ spin via two-qubit gates realized by microwave pulses and a final green optical pulse. As per injection, steps (2) and (3) must be completed within $\sim1 \ \mathrm{\mu s}$ of the time that the $^{14}$N$_\mathrm{S}$ nuclear spin is prepared in (1). However, this is not the most restrictive time. The most restrictive time is $\sim$180 ns for transport and capture that is set by the intrinsic spin dephasing time of diamond. As discussed previously, the capture rate is proportional to the electron probability density at the $^{14}$N$_\mathrm{S}$. To achieve a capture time of $\sim100$ ns, this probability density must be $\sim50 \ \mathrm{\mu m}^{-3}$, which roughly corresponds to confining the electron to a cube with sidelengths $\sim270$ nm. Therefore, the transport of the electron must occur over $<80$ ns and spatially controlled to this degree to achieve the capture within the total 180 ns. This significant challenge will be addressed in the next section.

Another important question that we have yet to address is whether it is possible for a free electron, other than the one prepared and injected, to be captured in error. In high purity diamond, such free electrons can only come from background N$_\mathrm{S}$ impurities. If the N$_\mathrm{S}$ impurity of the diamond is comparable to or better than the attainable $\sim10 \ \mathrm{\mu m}^{-3}$ \cite{isberg13}, then given the $\sim1.7$ eV depth of the N$_\mathrm{S}$ donor level, Fermi-Dirac statistics imply that there are essentially $\sim0$ free electrons thermally excited to the conduction band. It also means that there will be $<0.25$ free electrons injected by photoionization during the optical pulses (assuming a diffraction limited spot of $\sim300$ nm in diameter). Hence, we surmise that the chance of capturing a spurious electron, instead of the one that was prepared and injected, can be made negligibly small.

\section{Spin transport}

When the injected electron occupies a Bloch spin-orbital near the conduction band minimum, its spin is decoupled from its orbital degrees of freedom and evolves according to the pure Zeeman spin-Hamiltonian $H_\mathrm{cond.}=\gamma_es_zB$. As described previously, this decoupling is due to the absence of spin-orbit interaction in the $\Delta_1$ conduction band. Furthermore, the electron g-factor in the $\Delta_1$ band is close to the free electron value (within $\sim10^{-3}$) because the interband spin-orbit interaction is small ($\sim6$ meV) compared with the interband energy separation ($\sim5$ eV) \cite{dresselhaus}. Since $H_\mathrm{cond.}$ is identical to (\ref{eq:dopantspinhamiltoniansecular}) (when the $^{14}$N$_\mathrm{S}$ nuclear spin is correctly prepared), the electron will not accumulate a relative phase during its time in the Bloch spin-orbital. Consequently, in the absence of impurities and magnetic field inhomogeneities, spin dephasing during transport is limited to the intrinsic Elliot-Yafet spin relaxation mechanism of diamond.

The Elliot-Yafet mechanism is a mixture of interband spin-orbit interaction and momentum scattering by phonons \cite{restrepo12,zutic04}. The mechanism is dominant in solids with inversion symmetry and is particularly weak in diamond due to its small spin-orbit interaction, weak momentum scattering and large band separations. \textit{Ab initio} calculations predict that the Elliot-Yafet relaxation time in diamond at room temperature is the longest of any solid $T_2\sim T_1\sim180$ ns \cite{restrepo12}. As discussed in the last section, this defines the maximum time allowed for transport and capture and we have allocated $\sim80$ ns and $\sim100$ ns, respectively.

Given the decoupling of the injected electron's orbital and spin degrees of freedom, only its orbital degrees of freedom need to be considered in a model of its transport. For weak applied electric fields $\vec{E}_\mathrm{ap.}$ where transport occurs over timescales much longer than momentum and energy relaxation within the conduction band ($\sim1$ ps), a drift-diffusion model of the injected electron's probability density $\rho(\vec{r},t)$ may be adopted \cite{isberg13,ridley}
\begin{eqnarray}
\frac{\partial \rho}{\partial t} &=& \vec{\nabla}\cdot(D_n\vec{\nabla}\rho+\mu_n\rho\vec{E}) \nonumber \\
&& +k_I(t)N_I\delta(\vec{r}-\vec{r}_I)-\rho k_C \left[(1-N_C)\delta(\vec{r}-\vec{r}_C)\right. \nonumber \\
&&\left.+(1-N_I)\delta(\vec{r}-\vec{r}_I)\right], \nonumber \\
\frac{d N_I}{dt} & = & -k_I(t)N_I+\rho k_C(1-N_I), \nonumber \\
\frac{d N_C}{dt} & = & \rho k_C(1-N_C),
\end{eqnarray}
where $\mu_n=450 \ \mathrm{\mu m}^2/\mathrm{Vns}$ is the room temperature electron mobility in high-purity diamond \cite{isberg02}, $D_n=\mu_n k_BT/e=11 \ \mathrm{\mu m}^2/\mathrm{ns}$ is the room temperature electron diffusion constant, $e$ is the fundamental charge, $k_I(t)$ is the photoionization rate of the injecting $^{14}$N$_\mathrm{S}$ center located at $\vec{r}_I$ during the optical injection pulse, $k_C=\sigma_\mathrm{cap.}\sqrt{k_BT/m}$ is the (re-)capture rate per unit electron density of the (injecting)capturing $^{14}$N$_\mathrm{S}$ center located at ($\vec{r}_I$)$\vec{r}_C$, $N_I$ and $N_C$ are the probabilities that the electron is occupying the injecting and capturing centers, respectively,
\begin{eqnarray}
\vec{E} = \vec{E}_\mathrm{ap.}-\frac{e}{4\pi\epsilon}\vec{\nabla}\left[\frac{1-N_I}{|\vec{r}-\vec{r}_I|}+\frac{1-N_C}{|\vec{r}-\vec{r}_C|}\right],
\end{eqnarray}
is the total electric field including contributions from the charged centers, and $\epsilon$ is the electric permittivity of diamond.

\begin{figure}[hbtp]
\begin{center}
\mbox{
\subfigure[]{\includegraphics[width=0.9\columnwidth] {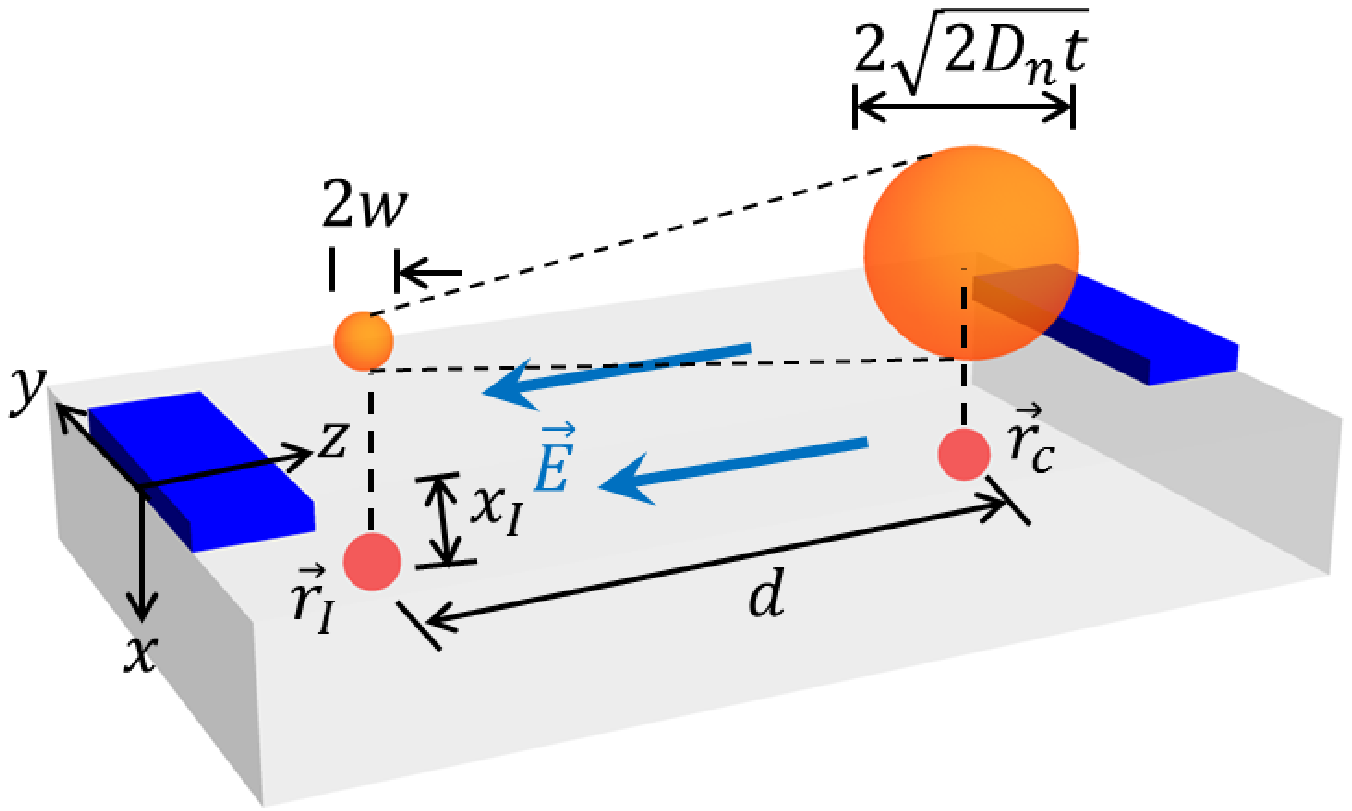}}
}
\mbox{
\subfigure[]{\includegraphics[width=0.9\columnwidth] {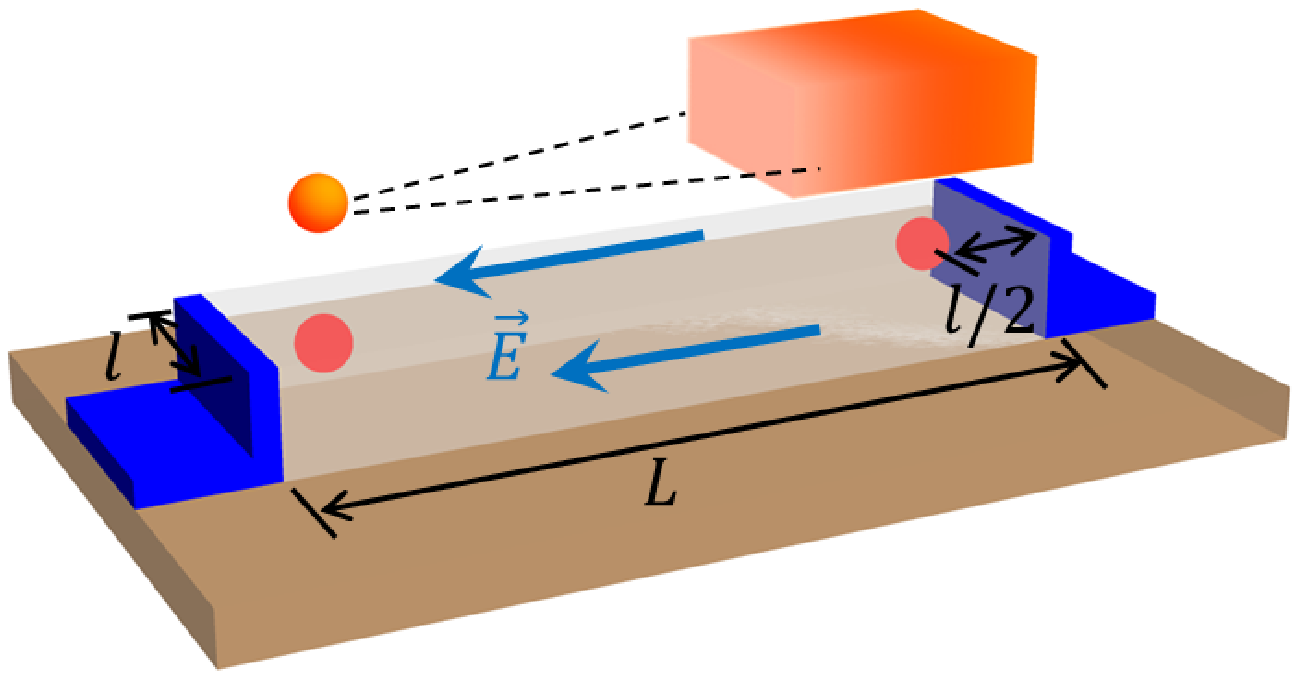}}
}
\mbox{
\subfigure[]{\includegraphics[width=0.9\columnwidth] {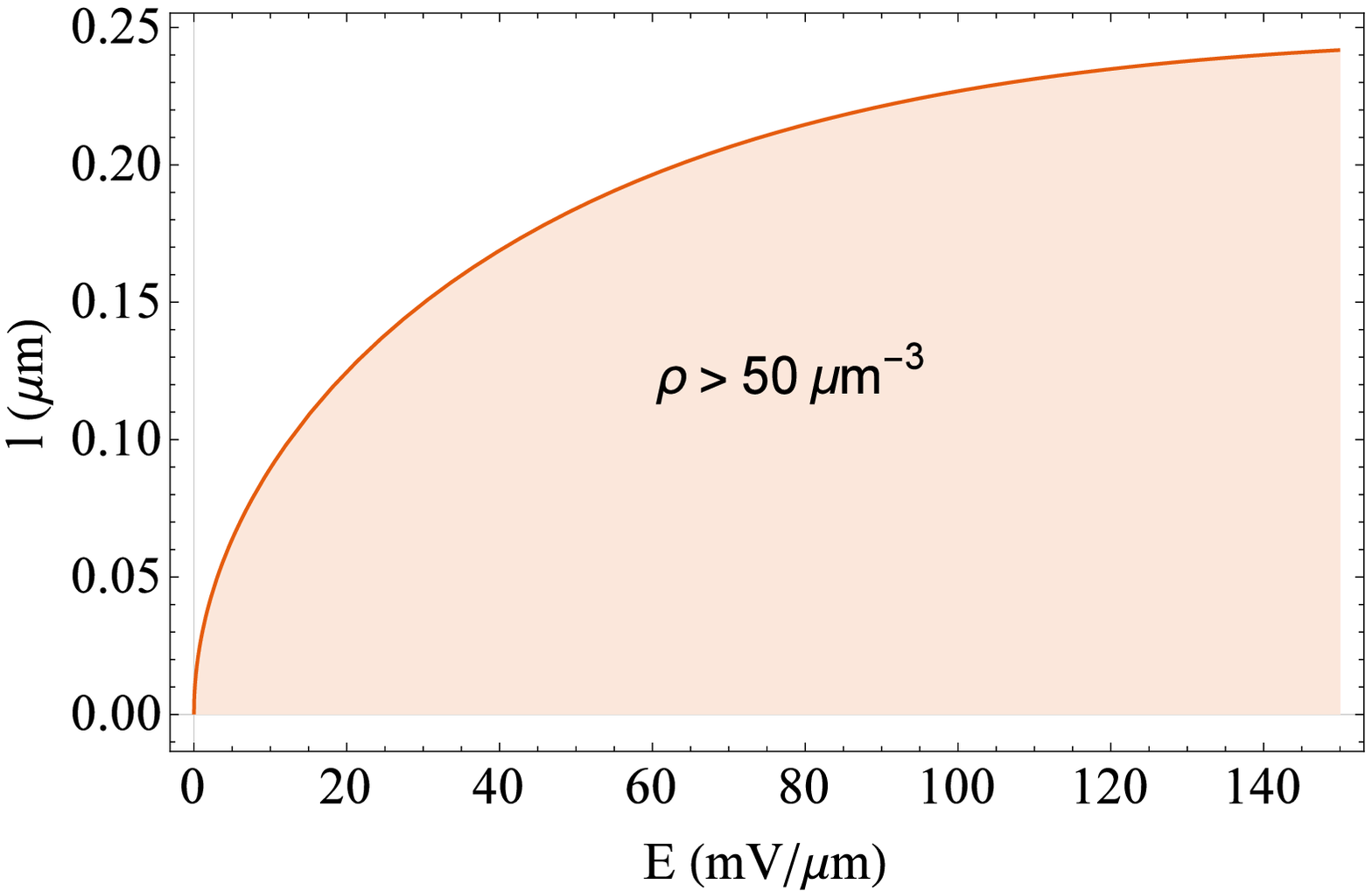}}
}
\caption{(color online)(a) A simple transport architecture consisting of two surface electrodes (dark blue) on bulk diamond (gray). (b) Example architecture for transport via a nanowire (gray) on top of a substrate (brown) with end electrodes (dark blue). In (a) and (b), the injection and capture centers are depicted as red spheres and the initial and final electron probability densities are depicted above in orange. All other labels are as described in the text. (c) The shaded region defines the nanowire dimension $l$ and applied electric field $E$ where the electron probability density at the capture center is the required $\rho>50 \ \mathrm{\mu m}^{-3}$ to yield a capture time $<100$ ns.}
\label{fig:transport}
\end{center}
\end{figure}

Consider the simple transport architecture depicted in Fig. \ref{fig:transport} that consists of two surface electrodes on bulk diamond. For simplicity, we will ignore the finite injection time and slow capture processes and make the approximation $\vec{E}\approx\vec{E}_\mathrm{ap.}\approx -E\hat{z}$ \cite{modelnote1}. In this case, the drift-diffusion model reduces to $\partial \rho/\partial t = D_n\nabla^2\rho-\mu_nE\partial\rho/\partial z$. The initial condition can be defined by approximating the injected electron probability density by a Gaussian $\rho(\vec{r},0)= e^{-|\vec{r}-\vec{r}_I|^2/2w^2}/\sqrt{2\pi w^2}$ centered at $\vec{r}_I=x_I\hat{x}$ and with width $w$. This approximate model has the separable solution $\rho = X(x,t)Y(y,t)Z(z,t)$, where
\begin{eqnarray}
X(x,t) & = & \frac{1}{\sqrt{4\pi D_nt^\ast}}\left[e^{-\frac{(x+x_I)^2}{4D_nt^\ast}}+e^{-\frac{(x-x_I)^2}{4D_nt^\ast}}\right], \nonumber \\
Y(y,t) & = & \frac{1}{\sqrt{4\pi D_nt^\ast}}e^{-\frac{y^2}{4D_nt^\ast}}, \nonumber \\
Z(z,t) & = & \frac{1}{\sqrt{4\pi D_nt^\ast}}e^{-\frac{z^2}{4D_nt^\ast}}e^{\frac{\mu_n E}{4D}(2z-\mu_nEt)},
\end{eqnarray}
$t^\ast=t+w^2/2D_n$, and the zero flux $(D_n\vec{\nabla}\rho+\mu_n\rho\vec{E})\cdot\hat{n}=0$ boundary condition has been enforced at the surface with unit normal vector $\hat{n}$. For $\sqrt{2D_nt}\ll x_I$, the mean position of $\rho$ is $\langle \vec{r}\rangle\approx x_I\hat{x}+\mu_n E t\hat{z}$ and its width $[\langle (x-\langle x\rangle)^2\rangle]^{1/2}=[\langle (y-\langle y\rangle)^2\rangle]^{1/2}=[\langle (z-\langle z\rangle)^2\rangle]^{1/2}\approx\sqrt{2D_nt^\ast}$ is isotropic and, for $D_n t\gg w^2$, approximately equal to $\sqrt{2D_nt}$. So, on average, the electron has been transported a distance $\mu_n E t$ and freely diffused over a volume with radius $\sqrt{2D_n t}$. Drawing upon the conclusions of the last subsection, to achieve capture within the allocated $\sim100$ ns, $\rho$ must be centered at $\vec{r}_C$ (i.e. $\mu_n Et=d$, where $\vec{r}_C=\vec{r}_I+d\hat{z}$) and have a radius $\sqrt{2D_n t}<135$ nm for a period of time $\sim100$ ns. However, after just $\sim1$ ps, the radius will be $>135$ nm \cite{modelnote2}, and so we can immediately conclude that this simple transport architecture cannot meet the requirements for capture because it does not adequately confine the diffusion of the electron probability density.

We can conceive of two strategies to confine the electron diffusion: (1) apply confining electric fields via additional surface electrodes  or (2) employ a diamond nanowire. There are two approaches to strategy (1): (A) use transverse electrodes to tightly focus the electron diffusion in the transverse directions (see Fig. \ref{fig:concept} for example) to achieve very high electron density at the capture center or (B) trap the electron using an electrostatic trap (e.g. a quadrupole trap \cite{iontrap}) and move the trap site from the injection to capture centers over the allocated $\sim80$ ns. A significant undertaking of device design is clearly required to properly assess strategy (1). Here, we leave this undertaking for future work and instead focus on strategy (2).

Consider the rectangular diamond nanowire with dimension $l\times l\times L$ sketched in Fig. \ref{fig:transport}. The side and end surfaces of the wire confine the electron probability density, such that after sufficient time $t\gg l^2/D_n \ \mathrm{and} \ L/\mu_n E$, it reaches the static distribution \begin{eqnarray}
\rho(\vec{r}) & \approx &\frac{\mu_n E}{D_n l^2} e^{-\frac{\mu_n E}{D_n}(L-z)},
\end{eqnarray}
where it has been assumed that $\mu_n EL/D_n\gg1$. $\rho$ is thus uniformly distributed over the transverse wire cross-section and confined to the wire endface by an exponential distribution with decay constant $\mu_n E/D_n$. Defining the distance from the capture center to the endface to be $(L-z_C)=l/2$, Fig. \ref{fig:transport}(c) plots the region in the parameter space $\{E,l\}$ where the electron probability density $\rho>50 \ \mathrm{\mu m}^{-3}$ is sufficiently large to yield a capture time $<100$ ns. For $l=0.2 \ \mathrm{\mu m}$, the applied electric field must be $E>63 \ \mathrm{mV}/\mathrm{\mu m}$, which is achievable \cite{bassett11}. In such electric fields, the center of the electron density is transported at a speed $\mu_nE>28 \ \mathrm{\mu m/ns}$ and can, in principle, be transported $>2$ mm during the allocated $\sim80$ ns. This is clearly more than adequate for a scalable device. Given that the technology already exists to fabricate diamond nanowires of these dimensions that contain NV and N$_\mathrm{S}$ centers \cite{babinec10,maletinsky12,momenzadeh14,andrich14}, diamond nanowires appear to be a very promising strategy to control spin transport.

This encouraging conclusion is, however, conditional on two assumptions that require further investigation: (1) the zero flux boundary condition is strictly enforced at the diamond surface so that there is perfect confinement of the electron to the nanowire's interior and, (2) spin relaxation during transport has not been altered by the structure and surfaces of the nanowire. Partial failure of these assumptions will result in electron losses and increased spin relaxation, respectively, and reduce the maximum attainable transport distance below the ideal estimate of $2$ mm.

Assumption (1) pertains to the termination and quality of the nanowire surfaces. Surface defects and impurities can act as electron traps and so must be avoided.  Nanowire surfaces exposed to air or vacuum must have a positive electron affinity in order to avoid loss of the electron. Oxygen and nitrogen terminated diamond surfaces are known to have positive electron affinities \cite{stacey15}. The latter, in particular, is predicted to have low densities of paramagnetic defects and therefore is more suited to spin-related applications \cite{stacey15}. Alternatively, it is possible that surface emission can be avoided by embedding the nanowire in another semiconductor or a dielectric, whose electron affinity is much smaller than diamond's. To assess the validity of assumption (1), future work should explore the yet to be studied electronic properties of diamond nanowires and their associated surface effects.

Assumption (2) pertains to the increase of spin relaxation by paramagnetic surface defects or the emergence of another transport relaxation mechanism other than the Elliot-Yafet mechanism. Whilst the former can, in principle, be controlled by fabrication precision and choice of surface termination, the latter is a more fundamental question. If the geometry of the nanowire sufficiently perturbs the inversion symmetry properties of the conduction band Bloch orbitals, then the D'yakonov-Perel' spin relaxation mechanism may emerge \cite{zutic04}. This is the dominant mechanism in solids without inversion symmetry \cite{zutic04}. Future \textit{ab initio} studies are needed to properly assess the emergence of the D'yakonov-Perel' mechanism in diamond nanowires.

\section{A room-temperature spin quantum bus}

\begin{figure}[hbtp]
\begin{center}
\mbox{
\subfigure[]{\includegraphics[width=0.9\columnwidth] {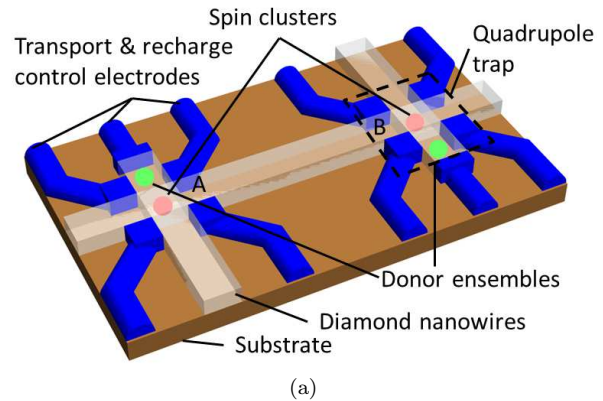}}
}
\mbox{
\subfigure[]{\includegraphics[width=1.0\columnwidth] {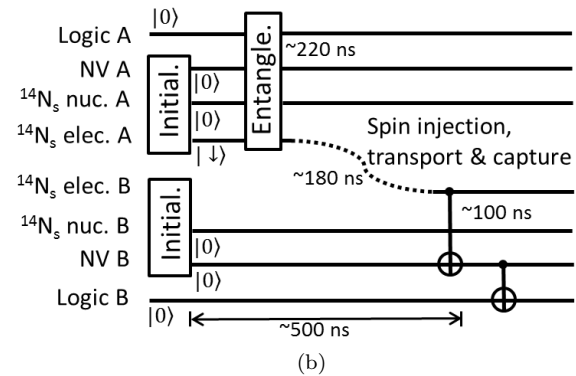}}
}
\caption{(color online)(a) Sketch of two nodes of the proposed spin quantum bus network constructed from diamond nanowires (white) on top of a substrate (brown). Spin clusters A and B (red) are located at the nodes and contain logic, NV and $^{14}$N$_\mathrm{S}$ spins. Nearby ensembles of N$_\mathrm{S}$ centers (green) for recharging the $^{14}$N$_\mathrm{S}$ centers in the clusters are located in compartments next to the nodes. Surface electrodes (blue) are used to control the spin transport between clusters and the recharge process. At each node, four of the electrodes are arranged in a quadruple configuration to aid in trapping the transport electron at the node. For clarity, optical, microwave and magnetic control structures (refer to Fig. \ref{fig:concept}) have not been included in the sketch. (b) The proposed protocol to generate entanglement between the logic qubits located in clusters A and B. See text for details of the protocol.}
\label{fig:spinbus}
\end{center}
\end{figure}

Based upon the conclusions of the previous sections, we propose that a room-temperature on-chip spin quantum bus can be constructed in diamond by connecting two defect spin clusters by a diamond nanowire [see Fig. \ref{fig:spinbus}(a)]. Each spin cluster contains at least a NV center, a $^{14}$N$_\mathrm{S}$ donor center and a logic spin qubit (either the N nuclear spin of the NV center, a $^{13}$C isotopic impurity or another NV center). The NV center acts as a fast local bus to prepare, control and readout the logic and $^{14}$N$_\mathrm{S}$ spins with high fidelity. The logic spin is chosen to have a long coherence time to minimize errors. The $^{14}$N$_\mathrm{S}$ center is included to enable optical spin injection, transport and detection between clusters as outlined in previous sections.  In order to recharge the $^{14}$N$_\mathrm{S}$ center on demand, each spin cluster is accompanied by a nearby ensemble of N$_\mathrm{S}$ centers to act as donors upon photoionization. Surface electrodes are used to control the spin transport between the clusters and the recharge process.

In a scalable design, similar to the one outlined in Ref. \cite{hollenberg06}, the spin clusters are located at the nodes of a network of diamond nanowires. Pursuant to the discussion of the last section, the surface electrodes must also be used to apply an electrostatic trap [eg. via a quadrupole configuration as per Fig. \ref{fig:spinbus}(a)] at the node in order to confine the transport electron and achieve the required capture rate. It is possible that there are other network designs where the spin clusters are not located at nodes, but somewhere where the nanostructured diamond can offer confinement of the transport electron without the need for electrostatic traps.

In previous sections, we have already discussed how this design of a spin quantum bus can be used to coherently transmit a spin state between two spin clusters. The other important function of a quantum bus is to generate entanglement between distant logic qubits. In our design, entanglement is generated between two distant logic qubits, located in spin clusters A and B, via the coherent transport of an entangled spin and entanglement swapping operations. In Fig. \ref{fig:spinbus} we outline a protocol to generate a maximally entangled Bell state of logic qubits A and B. The protocol begins with the initialization of the NV and $^{14}$N$_\mathrm{S}$ centers in the two clusters as defined in sections IIIB and IIIC. The desired Bell state is generated between Logic A and $^{14}$N$_\mathrm{S}$ electron spin A via a fast entanglement gate mediated by NV A, such as CPHASE \cite{qip8,cphase}. Optical spin injection, transport and capture proceeds as defined in sections IIIB, IIIC and IV and transfers the transport electron from $^{14}$N$_\mathrm{S}$ A to $^{14}$N$_\mathrm{S}$ B. Finally, the entanglement is swapped from the transport electron at $^{14}$N$_\mathrm{S}$ B to Logic B via a series of CNOT gates and NV B.

The only complication that this entanglement protocol adds to the ones already discussed in sections IIIB, IIIC and IV is that the entanglement gate must also be performed within the $\sim1 \ \mathrm{\mu s}$ period extending from initialization of the $^{14}$N$_\mathrm{S}$ nuclear spins to the final operation on the transport electron [the CNOT gate with NV B in Fig. \ref{fig:spinbus}(b)]. Since the complete initialization sequence defined in section IIIB includes $\sim500$ ns after the initialization of the $^{14}$N$_\mathrm{S}$ nuclear spins, there is $\sim500$ ns left for the gate operations and spin transport in Fig. \ref{fig:spinbus}(b). Given the NV-$^{14}$N$_\mathrm{S}$ gate operation times and spin transport times already discussed, the entanglement gate must be completed within $\sim220$ ns. Assuming a NV-$^{14}$N$_\mathrm{S}$ coupling of $\sim10$ MHz, the entanglement gate can be completed within $\sim220$ ns as long as the NV-logic qubit coupling is also $>10$ MHz. This is attainable for $^{13}$C logic qubits \cite{qip4}. Hence, we conclude that, in light of all of our considerations, it is feasible to realize the entanglement generation protocol of Fig. \ref{fig:spinbus}(b).

\section{Conclusion}
In this paper, we have explored an alternate approach to realizing a room-temperature spin quantum bus in diamond, a major unsolved problem in the development of scalable diamond QIP devices. Our approach was motivated by the unique properties of diamond paramagnetic defects and the predictions of extreme spin transport in diamond. By reviewing the properties of the NV, N$_\mathrm{S}$ and P$_\mathrm{S}$ centers, we identified novel optical spin injection and detection mechanisms. After assessing the mechanisms, we concluded that NV-$^{14}$N$_\mathrm{S}$ pairs are the best option for coherent spin injection and detection. We explored spin transport in diamond by applying a drift-diffusion model of electron transport to simple architectures. It became clear that adequate confinement of the transport electron, so that the required capture rates can be achieved, is the principal problem that must be overcome in order to realize spin transport between discrete spin clusters. We identified that diamond nanowires are a promising solution and that they may be combined with other strategies, such as electrostatic traps, in scalable designs of networked quantum buses. We finally proposed how such nanowire quantum buses can coherently transmit spin states between and generate entanglement of distant spin clusters.

The main purpose of this paper is to stimulate investigation of this alternate approach. As we have highlighted throughout our discussion, there are many aspects that require further experimental and theoretical study. These aspects range from the fundamental properties of the defect centers to the electrical properties of diamond nanowires. As this future work progresses, we expect that the preliminary ideas that we have outlined here will substantially evolve.

\begin{acknowledgments}
This work was supported by the Australian Research Council under the Discovery Project scheme (DP120102232), the National Science Foundation (NSF-1314205) and a grant (No. M-ERA.NET-1/2015) from the Research Council of Lithuania. We wish to thank A. Stacey, J. Meijer, K. Ganesan, A. Greentree, L.C.L. Hollenberg and S. Prawer for many useful discussions.
\end{acknowledgments}


\begin{thebibliography}{99}
\bibitem{audnote} The estimate of 66\% probability is based upon the simple picture that there is equal chance that any one of the three electrons occupying the $e$ spin-orbitals in the $^3E$ optical excited state are ejected. This picture is only strictly true if there is equal chance that the NV center's final state after photionization is any of the levels ($^4A_2$, $^2A_2$, $^2E$, $^2A_1$) of the $a_1e^2$ configuration. These levels are expected to be split by energies ($\sim$0.5 eV) comparable to optical energies. Thus, the final state of the NV center will depend on the optical excitation wavelength. We have worked through the implications of this complication and found that it does not alter our conclusions about the utility of the two-phonon mechanism.
\bibitem{estimatenote} To estimate ${\cal F}_\mathrm{NV}(\lambda)$ we first extracted the NV$^-$ absorption spectrum from Ref. \onlinecite{davies74} and rescaled it to units of absorption rate per unit optical power by comparing it to the partial photoexcitation spectrum in Ref. \onlinecite{aslam13}. We then extracted the one-photon photoionization spectrum from Ref. \onlinecite{aslam13}, which was in the similar units of photoionization rate per unit optical power. We finally inserted this data into the expression for ${\cal F}_\mathrm{NV}(\lambda)$. This estimation method allowed rates in approximately the same optical conditions to be compared.
\bibitem{modelnote1} Here, we have made the simplifying approximation that the surface electrodes apply a uniform electric field $\vec{E}_\mathrm{ap.}=-E\hat{z}$ in the region where transport is occuring. We have assumed that $\vec{E}_\mathrm{ap.}$ is much larger than the monopole electric fields produced by the charged injection and capture centers, so that they can be ignored and $\vec{E}\approx\vec{E}_\mathrm{ap.}$. Since the capture rate $\sim10$ MHz is much slower than the rates of injection $\sim0.1-1$ GHz and drift-diffusion $>\mathrm{\mu m/ns}$, it can be safely ignored when solving for the electron drift-diffusion. As injection and drift-diffusion have similar rates, the initial condition we define to solve for the electron drift-diffusion is not a true initial condition (as would be the case if injection were much faster), but rather represents a time-average of the electron probability density during injection. This time-average is quantified by the initial width $w$.
\bibitem{modelnote2} Although this result is outside the applicable timescale for the drift-diffusion model, we still conclude that it implies that the simple architecture does not adequately confine the electron probability density to achieve the required capture rates.
\end{thebibliography}
\end{document}